\documentclass[a4paper,fleqn,usenatbib]{mnras}
\hypersetup{draft}
\pdfoutput=1
\usepackage[T1]{fontenc}
\usepackage{ae,aecompl}
\usepackage{graphicx}
\usepackage{amsmath}
\usepackage{amssymb}
\usepackage[varg]{txfonts}
\usepackage{tablefootnote}
\usepackage{CJK}
\usepackage{hyperref}

\newcommand{\xgb}{$XGBoost$}

\newcommand{\afe}{[$\alpha$/Fe]}
\newcommand{\logg}{$\log{g}$}

\newcommand{\vel}{km s$^{-1}$}
\newcommand{\teff}{T$_\mathrm{eff}$}

\newcommand{\gai}{\textit{Gaia}}

\title[Chemical Clocks]{The GALAH Survey: Chemical Clocks}

\author[M. R. Hayden et al.]{
Michael~R.~Hayden$^{1,2}$\thanks{E-mail: michael.hayden@sydney.edu.au},
Sanjib~Sharma$^{1,2}$,
Joss~Bland-Hawthorn$^{1,2}$,
Lorenzo~Spina$^{3,2,4}$,\newauthor
Sven~Buder$^{5,2}$,
Martin~Asplund$^{6}$,
Andrew~R.~Casey$^{3,2}$,
Gayandhi~M.~De~Silva$^{7,8}$,\newauthor
Valentina~{D'Orazi}$^{4,3}$,
Ken~C.~Freeman$^{5,2}$,
Janez~Kos$^{9}$,
Geraint~F.~Lewis$^{1}$,
Jane~Lin$^{5,2}$,\newauthor
Karin~Lind$^{10}$,
Sarah~L.~Martell$^{11,2}$,
Katharine~J.~Schlesinger$^{5}$,
Jeffrey~D.~Simpson$^{11,2}$,\newauthor
Daniel~B.~Zucker$^{12,8,2}$,
Toma\v{z}~Zwitter$^{9}$,
Boquan Chen$^{1,2}$,
Klemen \v{C}otar$^{9}$,
Diane Feuillet$^{13}$,\newauthor
Jonti Horner$^{14}$,
Meridith Joyce$^{5,2}$,
Thomas Nordlander$^{5,2}$,
Dennis Stello$^{11,2}$,\newauthor
Thor Tepper-Garcia$^{1,2,15}$,
\begin{CJK*}{UTF8}{gbsn}
Yuan-sen Ting (丁源森) $^{16,17,18,5}$,
Purmortal Wang (王梓先)$^{1,2}$,
\end{CJK*}\newauthor
Rob Wittenmyer$^{14}$
\\
$^{1}$Sydney Institute for Astronomy, School of Physics, A28, The University of Sydney, NSW 2006, Australia\\
$^{2}$Centre of Excellence for Astrophysics in Three Dimensions (ASTRO-3D), Australia\\
$^{3}$School of Physics and Astronomy, Monash University, Australia\\
$^{4}$Istituto Nazionale di Astrofisica, Osservatorio Astronomico di Padova, vicolo dell'Osservatorio 5, 35122, Padova, Italy\\
$^{5}$Research School of Astronomy \& Astrophysics, Australian National University, ACT 2611, Australia\\
$^{6}$Max Planck Institute  for Astrophysics, Karl-Schwarzschild-Str. 1, 85741 Garching, Germany\\
$^{7}$Australian Astronomical Optics, Faculty of Science and Engineering, Macquarie University, Macquarie Park, NSW 2113, Australia\\
$^{8}$Macquarie University Research Centre for Astronomy, Astrophysics \& Astrophotonics, Sydney, NSW 2109, Australia\\
$^{9}$Faculty of Mathematics and Physics, University of Ljubljana, Jadranska 19, 1000 Ljubljana, Slovenia\\
$^{10}$Department of Astronomy, Stockholm University, AlbaNova University Centre, SE-106 91 Stockholm, Sweden\\
$^{11}$School of Physics, UNSW, Sydney, NSW 2052, Australia\\
$^{12}$Department of Physics and Astronomy, Macquarie University, Sydney\\
$^{13}$Lund Observatory, Department of Astronomy and Theoretical Physics, Box 43, SE-221\,00 Lund, Sweden\\
$^{14}$Centre for Astrophysics, University of Southern Queensland, Toowoomba, QLD 4350, Australia\\
$^{15}$Centre for Integrated Sustainability Analysis, School of Physics, The University of Sydney, NSW 2006, Australia\\
$^{16}$Institute for Advanced Study, Princeton, NJ 08540, USA\\
$^{17}$Department of Astrophysical Sciences, Princeton University, Princeton, NJ 08544, USA\\
$^{18}$Observatories of the Carnegie Institution of Washington, 813 Santa Barbara Street, Pasadena, CA 91101, USA\\
}

\date{}
\pubyear{2020}
\begin{document}
\label{firstpage}
\pagerange{\pageref{firstpage}--\pageref{lastpage}}
\maketitle

\begin{abstract}
Previous studies have found that the elemental abundances of a star correlate directly with its age and metallicity. Using this knowledge, we derive ages for nearly 250,000 stars of the GALAH DR3 sample using only their overall metallicity and chemical abundances. Stellar ages are estimated via the machine learning algorithm $XGBoost$, using main sequence turnoff stars with precise ages as our input training set.  We find that the stellar ages for the bulk of the GALAH DR3 sample are accurate to 1-2 Gyr using this method. With these ages, we replicate many recent results on the age-kinematic trends of the nearby disk, including the age-velocity dispersion relationship of the solar neighborhood and the larger global velocity dispersion relations of the disk found using \gai{} and GALAH. The fact that chemical abundances alone can be used to determine a reliable age for a star have profound implications for the future study of the Galaxy as well as upcoming spectroscopic surveys. These results show that the chemical abundance variation at a given birth radius is quite small, and imply that strong chemical tagging of stars directly to birth clusters may prove difficult with our current elemental abundance precision. Our results highlight the need of spectroscopic surveys to deliver precision abundances for as many nucleosynthetic production sites as possible in order to estimate reliable ages for stars directly from their chemical abundances. Applying the methods outlined in this paper opens a new door into studies of the kinematic structure and evolution of the disk, as ages may potentially be estimated for a large fraction of stars in existing spectroscopic surveys. This would yield a sample of millions of stars with reliable age determinations, and allow precise constraints to be put on various kinematic processes in the disk, such as the efficiency and timescales of radial migration.
\end{abstract}



\clearpage
\section{Introduction}
How the Milky Way formed and its evolution through time is one of the critical questions facing astrophysics today. Galactic Archaeology \citep{freeman2002} is the midst of a revolution in its attempt to answer these questions with the advent of \gai{} \citep{gaia2016}, enabling a plethora of new discoveries about the structure, formation, and evolution of the Galaxy. Even now, however, the origin of various Galactic substructures and the relative importance of secular processes such as blurring and migration is still a matter of great debate (e.g., \citealt{rix2013,bland-hawthorn2016}). Stars and stellar populations are one of the primary tracers by which the Galaxy can be studied, as stars contain the chemical imprint of the gas from which they formed, allowing the evolutionary history of the Galaxy to be traced through time \citep{freeman2002}. However, our knowledge and understanding of the Milky Way has been hampered by the lack of large samples of stars for which reliable age estimates are available.

Traditionally, stellar ages can be determined only for small subsets of the Hertzspring-Russell (H-R) diagram. Isochrone matching can be used to determine ages for stars along the main sequence turn off (MSTO) and sub-giant branch, where there is a large separation between stars of different ages in the \teff{} vs. luminosity plane. Ages can also be determined from high quality studies of solar twins, where stellar parameters can be determined with much higher precision than other stellar types. Asteroseismology provides an additional avenue for age determination, particularly for giant stars, by providing accurate mass and radius determinations. However, this requires extremely accurate photometry and long baseline observations, and is generally restricted to specific areas of the sky like the \textit{Kepler} \citep{kepler2010} fields. In giants, the C/N ratio can also be used as a proxy for age, as the first dredge up along the giant branch is mass dependant (e.g., \citealt{masseron2015,martig2016,ness2016,casali2019}). This relation can be calibrated using asteroseismic observations. However, each of these methods is only able to provide ages for a relatively small number of stars, and require significant effort to provide those ages. None of these methods are therefore able to estimate ages for the full sample of stars that would be observed by a large-scale spectroscopic survey such as GALAH. 

Studies of the chemical abundances of nearby stars have found that there are clear age-\afe{} relations (e.g., \citealt{haywood2013,2014A&A...562A..71B,hayden2017}), and that the \afe{} ratio is a good proxy for the age of a star in \afe-enhanced populations. For younger stellar populations belonging to the thin disk this is no longer the case, as most stars have similar \afe{} abundances; therefore \afe{} ceases to be a direct measurement of stellar age for thin disk stellar populations. However, for the thin disk slow neutron capture process (s-process) abundances have been found to have a strong correlation with a stars age. In particular, the [Y/Mg] or [Y/Al] ratio has been observed to have a tight correlation with stellar age in studies of local MSTO stars \citep{2015A&A...579A..52N,spina2016,2016A&A...593A..65N,2017MNRAS.465L.109F,bedell2018,spina2018,lin2020}. Additionally, there have been differences observed between the age-abundance trends of the thin and thick disks, or as a function of metallicity (e.g., \citealt{2017MNRAS.465L.109F,2019A&A...622A..59T,2020A&A...640A..81N,casali2020,lin2020}), Still, these studies highlight the potential of using chemical abundances such as yttrium to directly determine the age of a star, and potentially estimate a birth radius. Several studies have attempted to estimate the birth radii of stars by comparing their age and metallicities with models of Galaxy evolution (e.g., \citealt{minchev2018,feltzing2019}), placing a star at a specific place and time in the Galaxy when it was born.

\begin{table}
\centering
\caption{Data selection criteria used to select stars in this paper.}
\label{datacuts}
\begin{tabular}{cc}
\hline
Parameter & Value \\
\hline
sp\_flag & 0 \\
abundance\_flag $\mathrm{[X/Fe]}$ & 0 \\
SNR & $>20$ \\
\teff{} & $<6200$ K \\
$\sigma_{\mathrm{T_{eff}}}$ & $<150$ K \\
Spectral Fit ${\chi}^2$ & $<4$ \\
$\mathrm{[Fe/H]}$ &  $-1.0 < \mathrm{[Fe/H]} < 0.5$ dex \\
\hline
\end{tabular}
\end{table}

\begin{figure}
\centering
\includegraphics[width=3.2in]{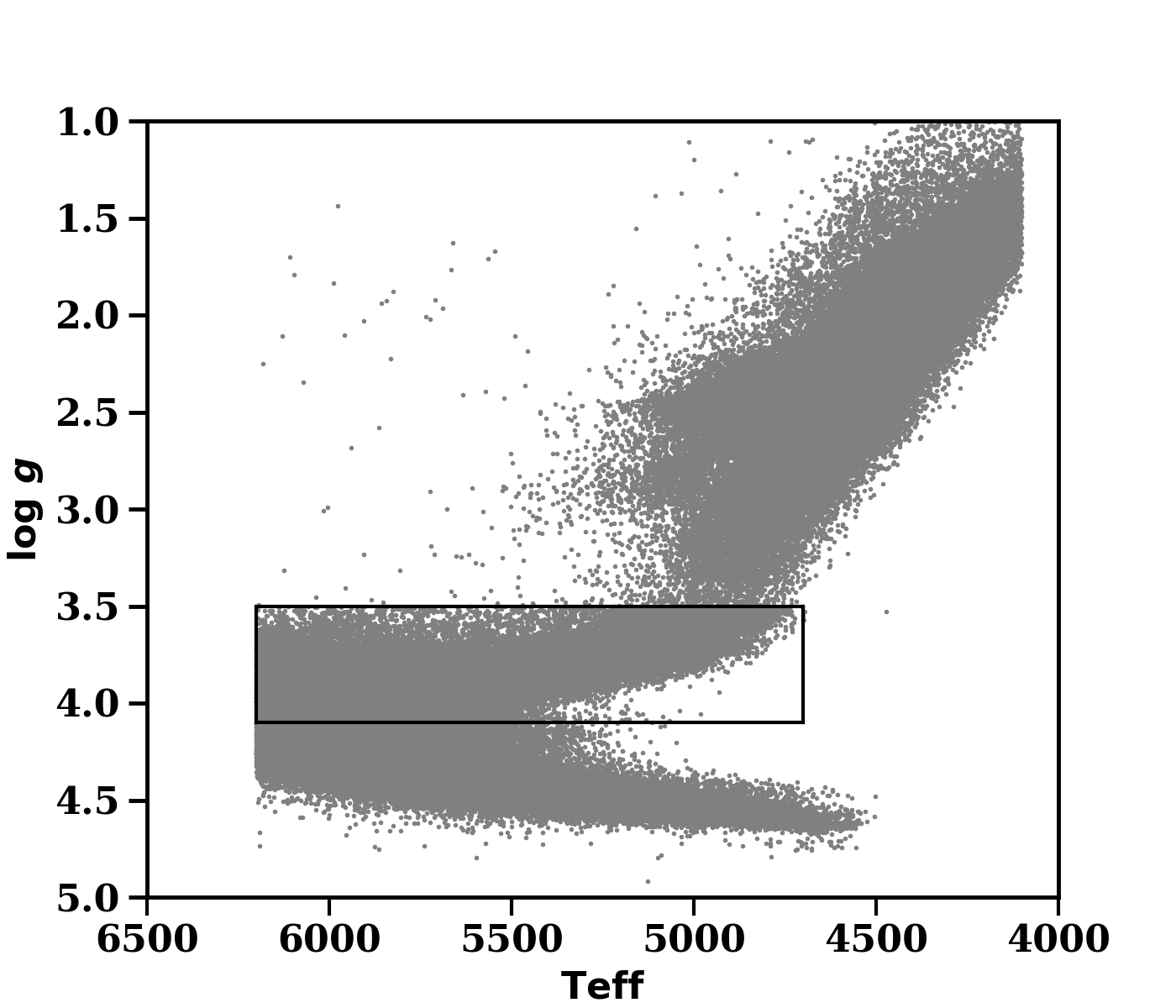}
\caption{The H-R diagram for the sample presented in this paper. The MSTO selection criteria is shown by the black box.}
\label{hrsample}
\end{figure}

Tagging a star to a particular birth location is known as chemical tagging \citep{freeman2002}. In strong chemical tagging, as outlined by \citet{2010ApJ...713..166B}, a star can be assigned to an individual birth cluster by measuring many abundances to high precision. Open clusters have been found to have very uniform chemical abundances, with scatter lower than 0.03 dex, which is the typical measurement uncertainty in such studies \citep{desilva2006,bovycluster2016}. However, it is unclear to what degree each cluster has a unique chemical signature, or how this changes with time. For example, if the thick disk formed from many massive clumps (\citealt{clarke2019}, but see \citealt{ting2016} also), these clumps may have unique chemical signatures enabling strong chemical tagging, while the smaller clusters forming in the thin disk today may not be very chemically distinct from other nearby star forming regions. If the ISM is in general well-mixed and the mixing timescale is short, it is possible that tagging to a specific cluster may be very difficult with the abundance precision in current spectroscopic surveys. Even in the case where strong chemical tagging is not possible, however, weak chemical tagging can still be extremely useful for Galactic Archaeology. Tagging a star to a general location in a Galaxy (i.e., a birth radius), rather than a specific star cluster, is known as weak chemical tagging. Weak chemical tagging allows the study of the spatial and kinematic structure of the Galaxy through time, and can aid in determining the effectiveness of secular processes such as blurring or migration \citep{sellwood2002,schonrich2009}. 

\begin{figure}
\centering
\includegraphics[width=3.2in]{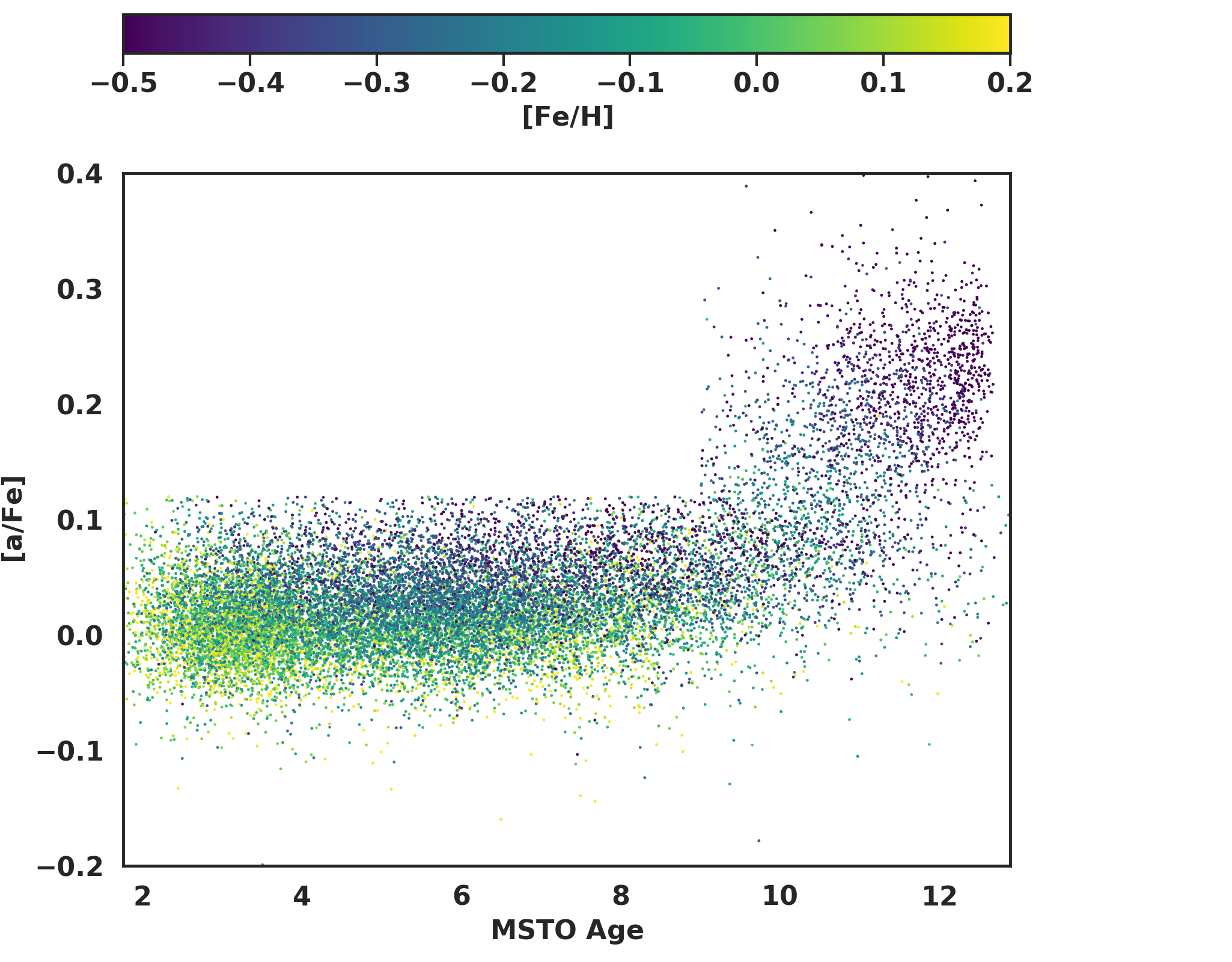}
\caption{The age-\afe{} relation for the MSTO training set.}
\label{agealpha}
\end{figure}

\begin{table}
\centering
\caption{Data cuts used on the GALAH DR3 sample to select training set of MSTO stars.}
\label{mstocut}
\begin{tabular}{cc}
\hline
Parameter & Value \\
\hline
\logg & $3.5<\log{g}<4.1$ \\
SNR & $>45$ \\
$\tau$ (Gyr) & $>1.75$ Gyr \\
$\frac{\sigma_{\tau}}{\tau}$ & $<0.2$\\
\hline
\end{tabular}
\end{table}

\citet{ness2019} used APOGEE \citep{majewski2017} data and found that the age and metallicity of a star alone could be used to estimate the initial orbits and abundances for stars, and that deviations from calculated age-abundance relations were quite small, within the measurement errors. These authors argue that this makes strong chemical tagging unrealistic for much of the disk, at least with current measurement precision. \citet{casali2020} use high resolution HARPS observations, and argue that the age-abundance relations, particularly those involving yttrium, are not universal, as they find that there are variations between the [Y/Mg]-age relation as a function of metallicity. These s-process elements were not observed by APOGEE.  \citet{casey2019} use GALAH \citep{desilva2015} to group stars by their chemical abundances, in particular by latent factors linked to the relative contribution of different nucleosynthetic production sites (e.g., \citealt{kobayashi2020}); these authors identify six different latent factors linked to elemental production sites are required to explain the abundance trends observed in GALAH. \citet{sharma2020c} also use GALAH data to measure age-abundance relations for nearly 30 elements and find that almost all of the elements studied show significant trends with age. They find, similar to \citet{ness2019}, that to first order age and metallicity alone can be used to estimate abundances across the disk, but do find minor deviations from these relations for some elements often related to the SNR or \teff. 

The arguments laid out in \citet{ness2019} and \citet{sharma2020c} have profound implications for the study of the kinematic structure and evolution of the disk. These authors argue that the age and metallicity alone can be used to predict chemical abundances. If correct, these relations can also be inverted: the age of a star can be inferred based on its overall metallicity and chemical abundances. In essence, the chemical abundances of a star act as a clock for the chemical evolution of the Galaxy, from which we can date the age of a star based on its abundances. This means that if abundances can be measured to high precision, ages can potentially be estimated for every star in a stellar survey, expanding sample sizes with reliable ages by orders of magnitude compared to existing studies. In the ideal case, a survey would measure abundances for as many different metal production sites as possible (Type II Supernovae, Type Ia Supernovae, s-process, and rapid neutron capture process). In practice, not all surveys are able to observe elements from all production sites given their resolution or wavelength coverage. Additionally, some elements may be difficult to measure for different stellar types or due to non-LTE effects. Compromises must be made when selecting which chemical abundances to use between having more abundances that make an age estimate more precise, or using fewer abundances that are well measured for a larger fraction of the stars in a stellar survey.  

In this paper, we estimate the ages of several hundred thousand stars directly from their chemical abundances as measured in GALAH DR33 \citep{buder2020}. We use MSTO stars as a training set for Bayesian and machine learning models for age estimation. With these ages, we reproduce recent age-kinematic observations of the disk with a sample size an order of magnitude larger than was previously available. GALAH DR3 provides abundance determinations for nearly 30 elements, allowing us to be selective and choose elements that are well estimated for a large fraction of the sample while also covering the different nucleosynthetic production sites in the Milky Way. 

This paper is organized as follows: in Section 2, we describe the sample selection, data cuts, and the MSTO training selection used in our analysis. In Section 3, we outline the methods we use to determine stellar ages for a large fraction of the GALAH DR3 sample, using Bayesian analysis as well as \xgb{} \citep{xgboost2016} to generate a model for the age of a star based on its chemical abundances. In Section 4, we demonstrate the reliability of our age determinations and reproduce many of the recent kinematic studies of the Galaxy. In Section 5, we discuss our findings in the context of strong and weak chemical tagging, implications for future kinematic studies of the disk, the chemical evolution of the Galaxy, and future survey design. 

\begin{figure}
\centering \includegraphics[width=0.49\textwidth]{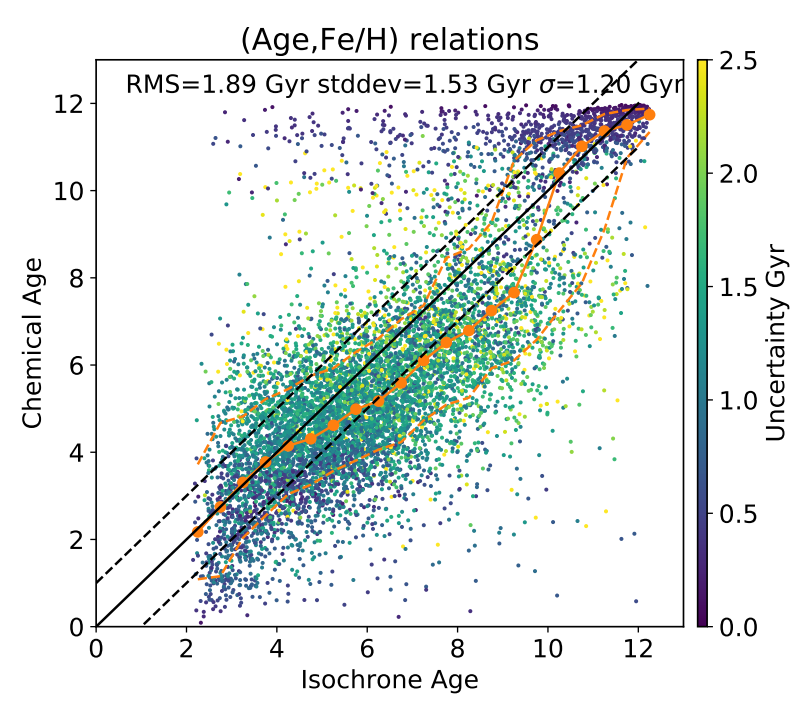}
\caption{Comparison of chemical ages with those based on isochrones and stellar parameters.
\label{fig:chemical_ages_ar1}}
\end{figure}

\section{Data}

Spectroscopic data are taken from GALAH DR3 \citep{buder2020}, and along with additional fields from the K2-HERMES \citep{2018AJ....155...84W} and \textit{TESS}-HERMES surveys \citep{sharma2018}. GALAH uses the High Efficiency and Resolution Multi-Element Spectrograph (HERMES, \citealt{2015JATIS...1c5002S}) instrument, which is a high resolution ($R\sim28 000$) multi-fibre spectrograph mounted on the 3.9 meter Anglo Australian Telescope (AAT). HERMES covers four wavelength ranges ($4 713-4 903${\AA}, $5 648-5 873${\AA}, $6 478-6 737${\AA}, and $7 585-7 887${\AA}), carefully selected to maximise the number of elemental abundances that are able to be measured. Observations are reduced through a standardised pipeline developed for the GALAH survey as described in \citet{kos2017}. Stellar atmospheric parameters and individual abundances are derived using SME \citep{valenti1996,piskunov2017}. The precision of individual abundances [X/Fe] is typically $\sim0.05$ dex, while the random errors in radial velocities are $\sim100\ \mathrm{m\ s^{-1}}$ \citep{zwitter2018}. In this analysis, we use a selection of 13 well measured elemental abundances from GALAH DR3: [Fe/H], [Mg/Fe], [Ca/Fe], [Ti I/Fe], [Si/Fe], [O/Fe], [Mn/Fe], [Cr/Fe], [Na/Fe], [K/Fe], [Y/Fe], [Ba/Fe], and [Sc/Fe] which span a range of nucleosynthetic production sites.

\begin{figure}
\centering
\includegraphics[width=3.2in]{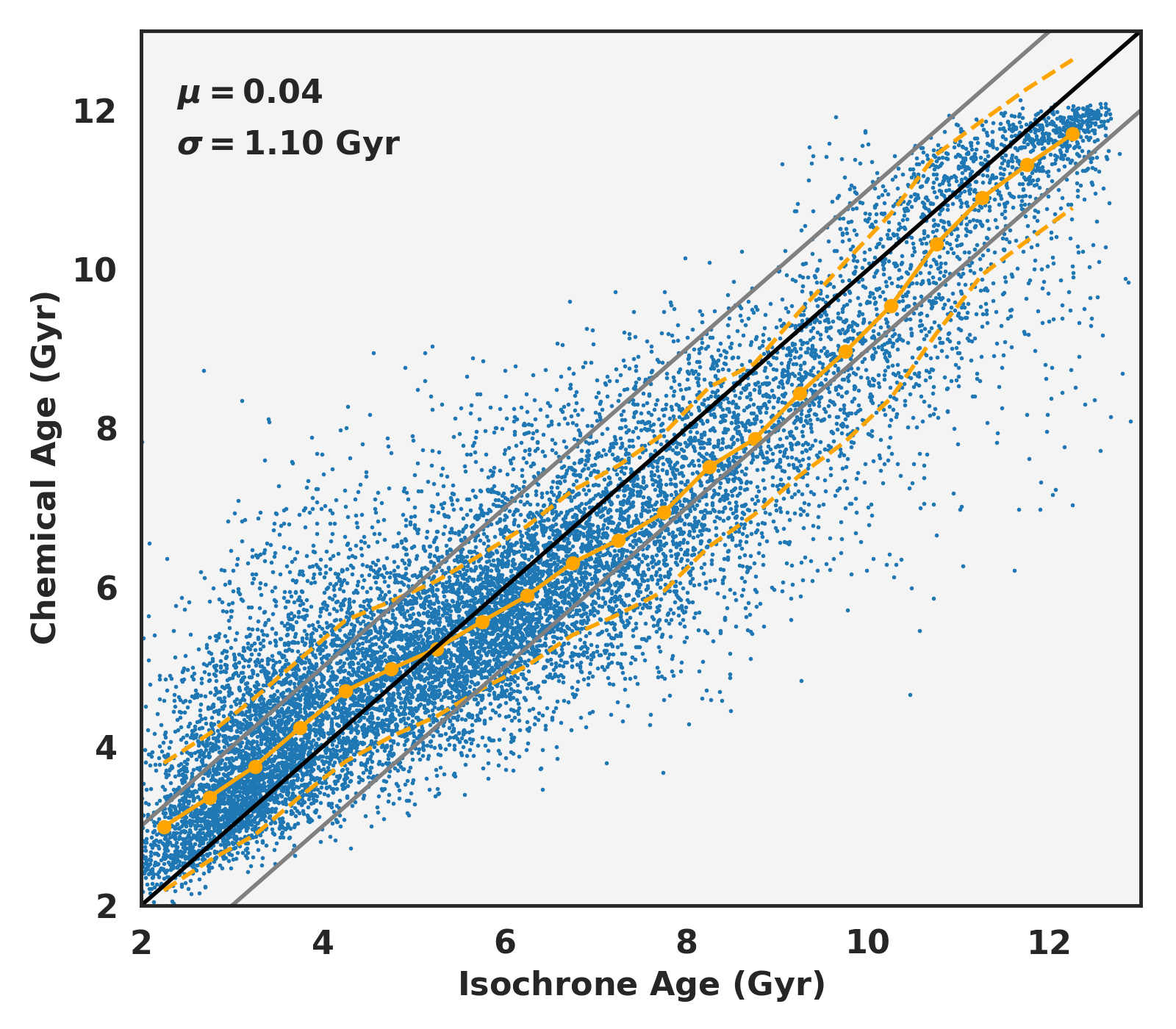}
\caption{The chemical age versus age determined from isochrone matching for the training set of MSTO stars using \xgb. The black line denotes a 1:1 correlation, while the grey lines denote $\pm1$ Gyr. The orange line denotes the median and one sigma scatter about the relation.}
\label{trainingset}
\end{figure}

\begin{table}
\centering
\caption{\xgb{} Parameters Used In Model Generation}
\label{xgboostparam}
\begin{tabular}{cc}
\hline
Parameter & Value \\
\hline
${\Gamma}$ & 5 \\
Max Depth of Tree & 7 \\
Subsample Ratio & 0.7 \\
Learning Rate $\eta$ & 0.05\\
Minimum Child Weight & 5 \\
Column Sampled by Tree & 0.8\\
\hline
\end{tabular}
\end{table}

Because of slight differences in both abundance precision and zeropoint shifts between giants ($\log{g}<3.5$) and dwarfs ($\log{g}>3.5$), we report results for these subsamples as well as all of GALAH DR3 separately in this paper; however the same model trained on MSTO stars is applied to both subsamples. We implement several quality cuts on the signal-to-noise ratio of the spectra, stellar parameter and abundance flags, as well as a temperature cutoff of \teff$<6200$ K, as outlined in Table \ref{datacuts}. We require a SNR$>20$, a quality flag$=0$ for both the stellar parameters and the individual abundances (see the GALAH DR3 release paper, \citealt{buder2020}). The reason for the \teff{} cut at 6200 K is we have found that there are significant issues for abundances measured for the hottest stars. We restrict our parameter space to that covered by the disk, $-1<$[Fe/H]$<0.5$, as the age-abundance relations of any accreted objects, e.g., the halo, follow a different enrichment history. These cuts give us a sample of 155,519 dwarf stars and 92,645 giant stars for which we determine ages for in this paper.

\begin{figure}
\centering
\includegraphics[width=3.2in]{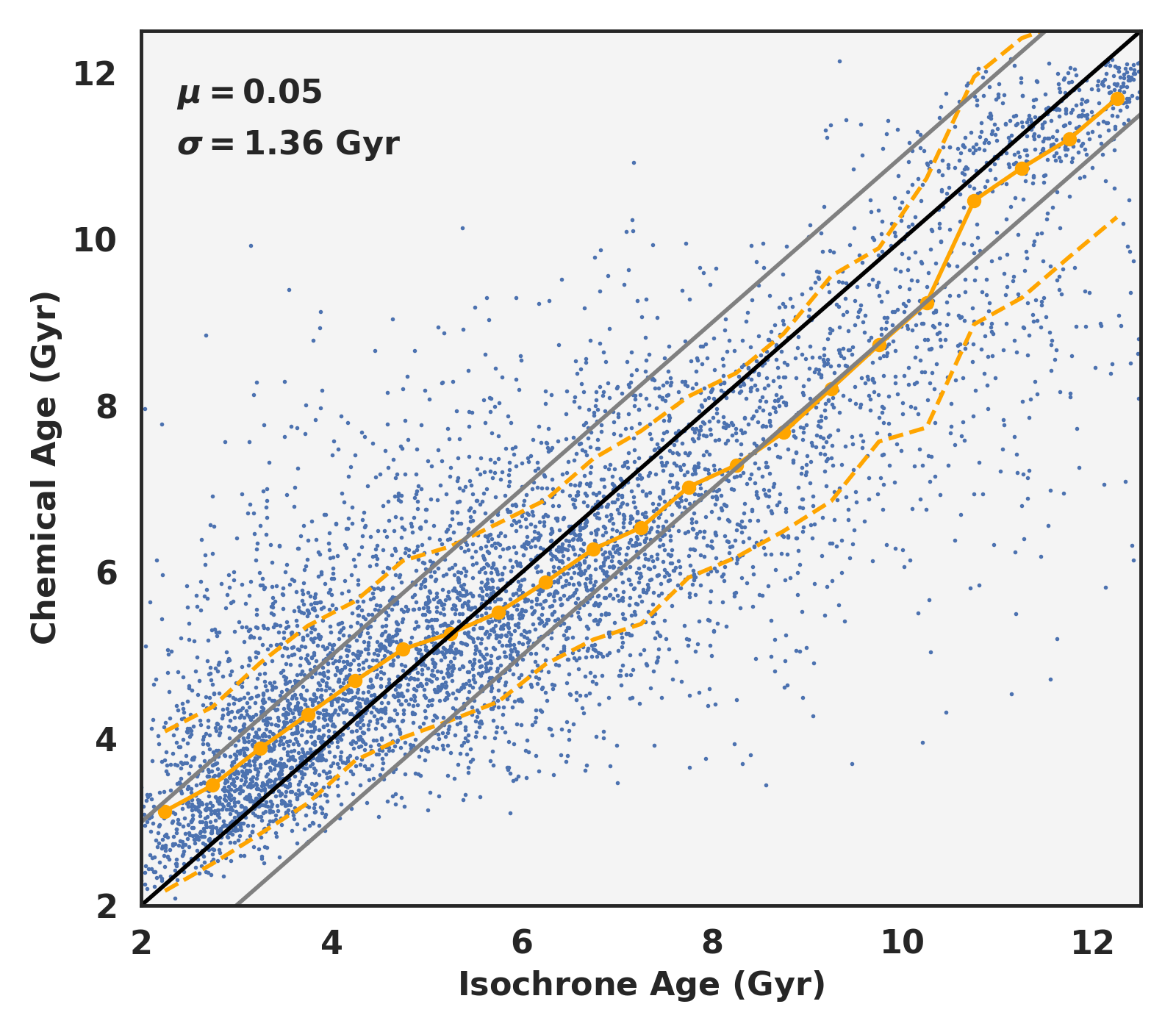}
\caption{The chemical age versus age determined from isochrone matching for the test set of MSTO stars using \xgb. The black line denotes a 1:1 correlation, while the grey lines denote $\pm1$ Gyr. The orange line denotes the median and one sigma scatter about the relation. Note that the test set performs worse than the training set as \xgb{} is slightly overfitting. However, even here the scatter is relatively small, and even the largest systematic age error is $<1$ Gyr for the worst performing cases (age between $8-10$ Gyr.}
\label{testset}
\end{figure}

For the training set we use main sequence turn off stars (MSTO), with ages computed with the code BSTEP \citep{sharma2018} with isochrone matching using PARSEC-COLIBRI isochrones \citep{marigo2017}. BSTEP provides a Bayesian estimate of intrinsic stellar parameters by comparing observed parameters to those of the PARSEC-COLIBRI isochrones. For determining the age of an MSTO star through BSTEP, the relevant stellar and astrometric parameters are  \teff, \logg, [Fe/H], \afe, $J$, $Ks$, and parallax. Astrometric parameters are taken from \textit{Gaia} DR2 \citep{gaia2018dr2,2018A&A...616A...2L}. The stellar distances and velocities used in this analysis are taken from the GALAH DR3 dynamics value added catalog, see \citet{buder2020} for details.

The relative accuracy for ages derived for MSTO stars in BSTEP is on the order of 10-15\%. With these high precision ages, MSTO stars make an ideal training set for age-abundance relations. In addition to the abundance and stellar parameter quality conditions applied to the rest of the GALAH sample, we apply additional criteria for the MSTO training set as outlined in Table \ref{mstocut}. To isolate the MSTO stars, we use a cut in \logg{} space with $3.5<\log{g}<4.1$, as well as stricter criteria on the SNR, with SNR$>45$ for the training set. See Fig. \ref{hrsample} for the H-R diagram of the sample which meets our selection criteria. The black box denotes the MSTO sample for which accurate ages are able to be estimated from isochrones. There are issues in determining the abundances for the youngest stars, $\tau<1.75$ Gyr, and we elect to remove these stars from the training set. For stars belonging to the thick disk (\afe$>0.15$), we require these stars to have an age $\tau>8$ Gyr (i.e., we remove the young $\alpha$-rich stars from the training set, see \citealt{chiappini2015}), see Fig. \ref{agealpha} for the age-$\alpha$ relation found in our training set. We also require the fractional uncertainty in the age determination from the isochrone matching to be better than 20\%, i.e., $\frac{\sigma_{\tau}}{\tau}<0.2$. These restrictions, along with the quality cuts from Table \ref{datacuts}, leave a sample of 15,424 MSTO stars belonging to the training set.

\section{Methods}

Ages can be estimated from chemical abundances in several ways. Initial studies of the [Y/Mg] ratio used linear fits (e.g., \citealt{2016A&A...593A..65N,bedell2018,spina2018}). However, variations have been found with metallcity in the age-abundance trends (e.g., \citealt{2017MNRAS.465L.109F,2019A&A...622A..59T,casali2020,lin2020}), and as our sample covers a large range in Galactocentric radii, more sophisticated methods are required. \citet{sharma2020c} determine age-abundance relations for various elements measured in GALAH DR3, under the assumption that an abundance can be determined given its age and metallicity. For elements which follow this assumption, this formalism can be inverted, and an age determined given a metallicity and a set of abundances, using Bayes Theorem. Let $X_i$ denote the observed abundance [X/Fe] of the $i$-th element with measurement uncertainty $\sigma_{Xi}$. Let $F$ be the observed metallicity [Fe/H] and $\sigma_F$ its uncertainty. Given age $\tau$ and metallicity $F'$ we can predict the abundance $X_{i,{\rm Model}}(\tau,F')$ using a model with some intrinsic dispersion $\epsilon_{Xi}$. Using Bayes theorem the probability distribution of age of a star given its metallicity and elemental abundances can then be written as: 
\begin{align}
p(\tau|X_i,\sigma_{Xi},F,\sigma_F)= \int {\mathrm{ d}} F' p(\tau|F')p(F'|F,\sigma_F) \times \\
    \prod_i p(X_i | \tau,F',\sigma_{Xi}).
\end{align}
where 
\begin{equation}
p(X_i | \tau, F', \sigma_{Xi}) = \mathcal{N}\left(X_i|X_{i,{\mathrm{ Model}}(\tau,F')},\sigma_{Xi}^2+\epsilon_{Xi}^2\right)
\end{equation}
and
\begin{equation}
p(F' | F,\sigma_{F})=\mathcal{N}\left(F|F',\sigma_{F}^2\right) 
\end{equation}

We apply this methodology with the age-abundance trends measured in \citet{sharma2020c}, and results for the age determination using Bayesian analysis are shown in Fig. \ref{fig:chemical_ages_ar1}. There is in general a good agreement between the age determined from isochrone matching and the age determined using Bayes theorem, although there are significant number of outliers. The primary benefit of the Bayesian approach is that the errors in abundances are directly incorporated into the age estimation, along with the fact that missing or flagged abundances still allow an age to easily to be determined.

\begin{figure}
\centering
\includegraphics[width=3.2in]{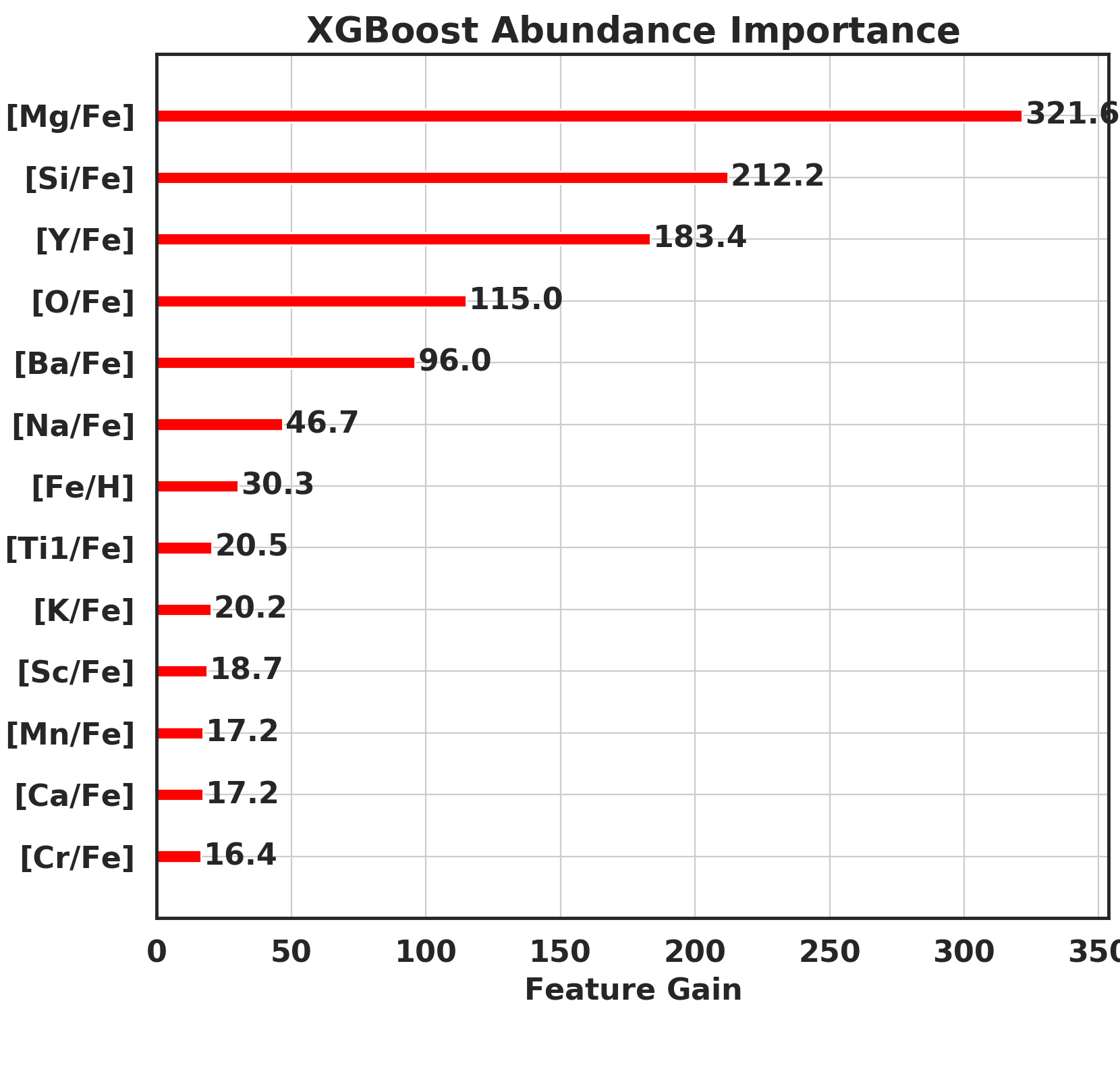}
\includegraphics[width=3.2in]{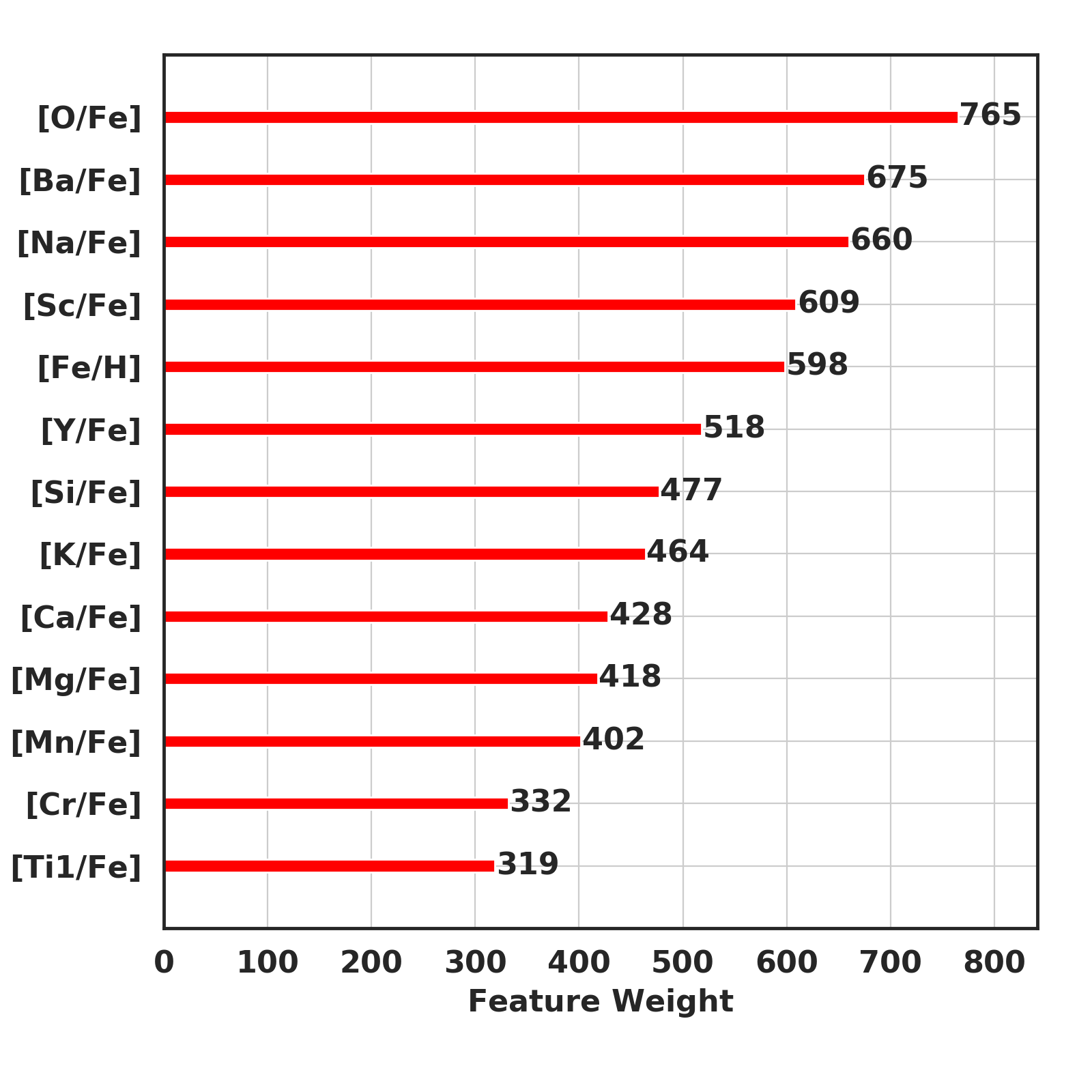}
\caption{\textbf{Top:} The relative gain of different chemical abundances in determining the age of a star. \textbf{Bottom:} The relative weight of different chemical abundances in determining the age of a star.}
\label{xgbgain}
\end{figure}

\begin{figure}
\centering
\includegraphics[width=3in]{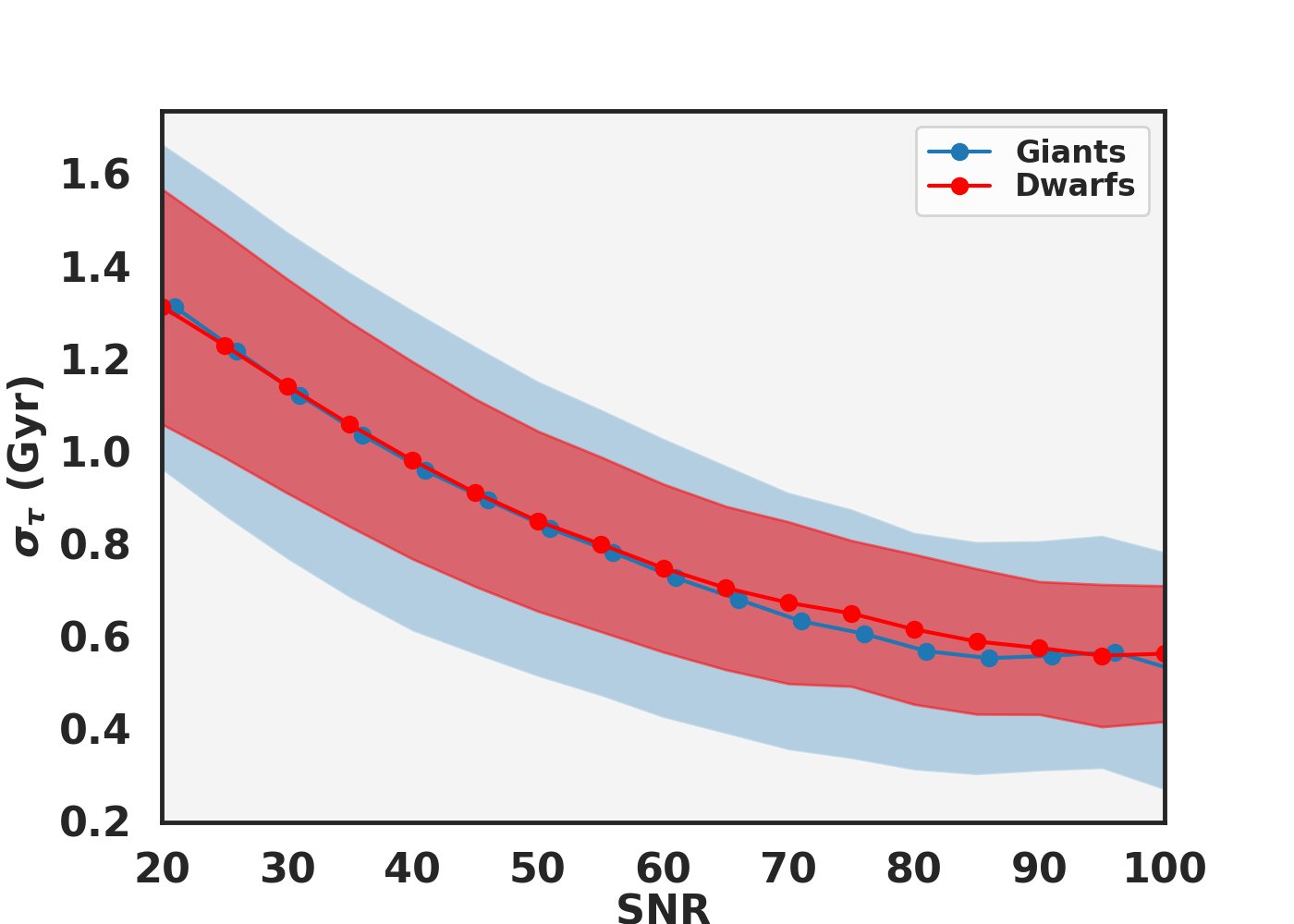}
\includegraphics[width=3in]{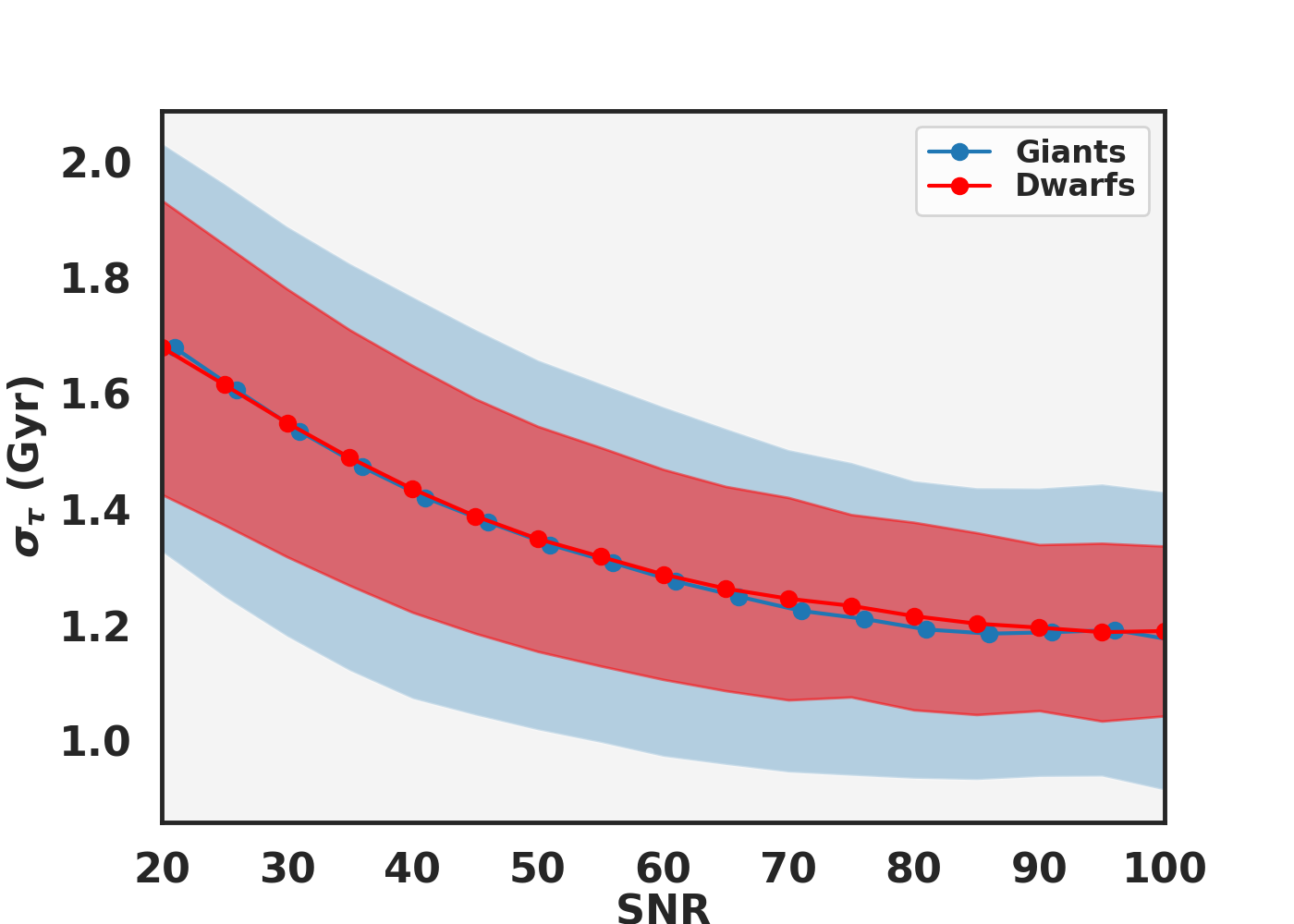}
\caption{\textbf{Top:} The random age errors due to abundance uncertainties as a function of SNR. The giants and dwarfs have similar median trends, but the giants have significantly more scatter as a function of SNR than dwarfs. \textbf{Bottom:} The total uncertainty in the age estimates, combining the random and systematic errors. }
\label{ageuncertainty}
\end{figure}

In addition to Bayesian analysis, ages can also be estimated using machine learning tools such as Neural Networks or gradient boosting algorithms. For this paper, we use the gradient boosting algorithm \xgb{} \citep{xgboost2016}. We split our initial sample of 15,424 MSTO stars into a training set (70\% of the original MSTO sample) and test set (30\% of the MSTO sample). The training set uses 13 chemical abundances described in the data section as input, along with the desired output parameter of age $\tau$ determined from MSTO isochrone fitting. We ran a 5-fold cross validation grid of several hundred thousand \xgb{} hyper parameters to obtain the model that best reproduced the trends of the test set while trying to minimize the over-fitting of the training set. The parameters used for \xgb{} are given below in Table \ref{xgboostparam}. 

The chemical age estimated from the best fit model relative to the input isochrone age determination is shown in Fig. \ref{trainingset} and \ref{testset}. The training set is slightly overfit relative to the test set, with a scatter $\sim1.1$ Gyr in the training set compared to the test set ($\sigma_{\tau}\sim1.35$ Gyr). However, this overfitting problem is less severe than in other methods we attempted (e.g., neural networks), while still giving age determinations for the test set that are more robust than those determined using Bayesian analysis. This model is able to reproduce the general age trend found in the test set, with a slight overestimate of ages for stars $\tau<6$ Gyr and a slight underestimate in ages for stars $\tau>6$ Gyr, i.e., the slope of the chemical age vs isochrone age relation is generally shallower at intermediate ages relative to the 1:1 line, as shown by the running median (orange line) in Fig. \ref{testset}. Most importantly, the number of critical failures/outliers is significantly reduced compared to the Bayesian age estimates.

\begin{figure*}
\centering
\includegraphics[width=3.2in]{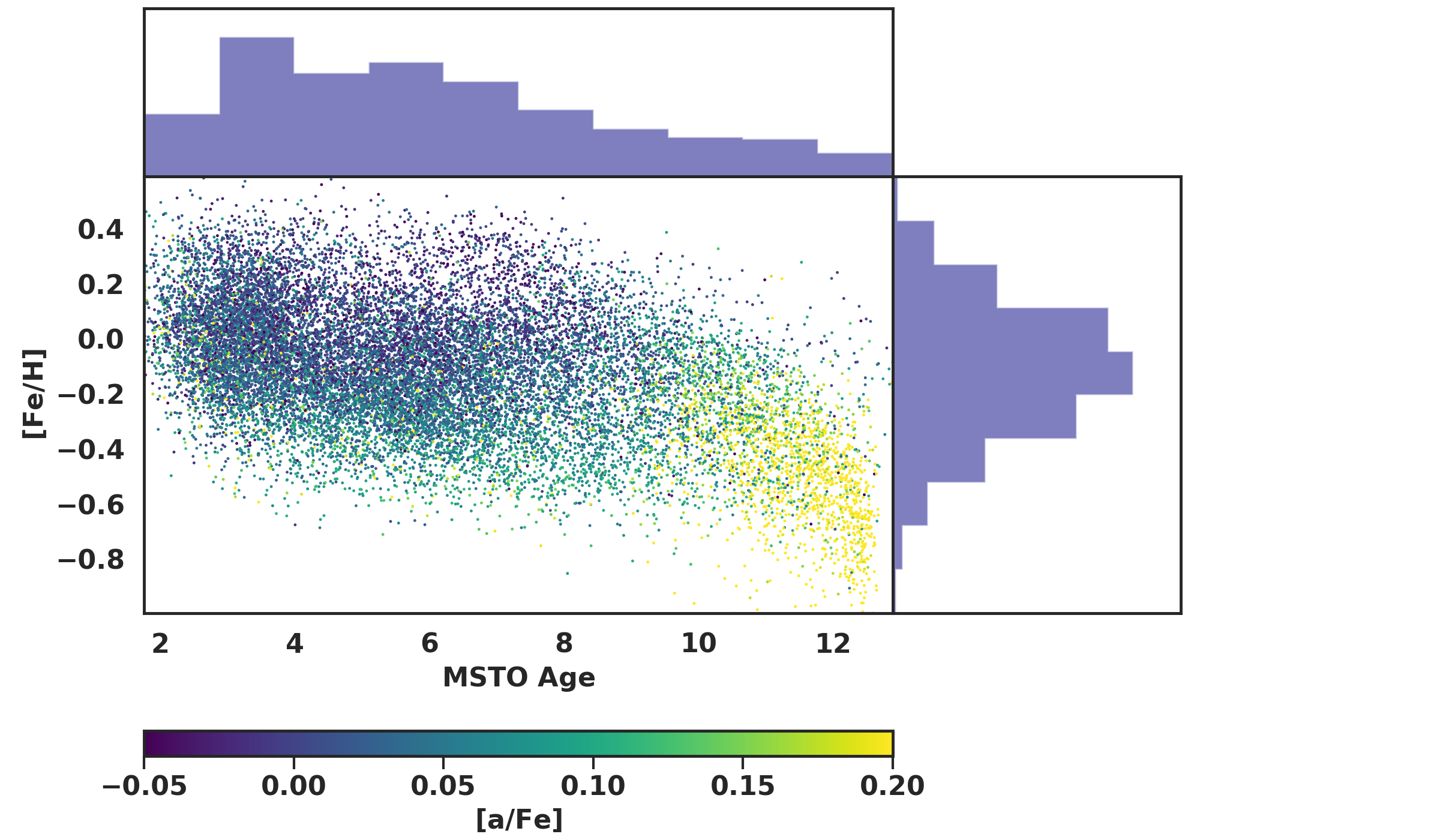}
\includegraphics[width=3.2in]{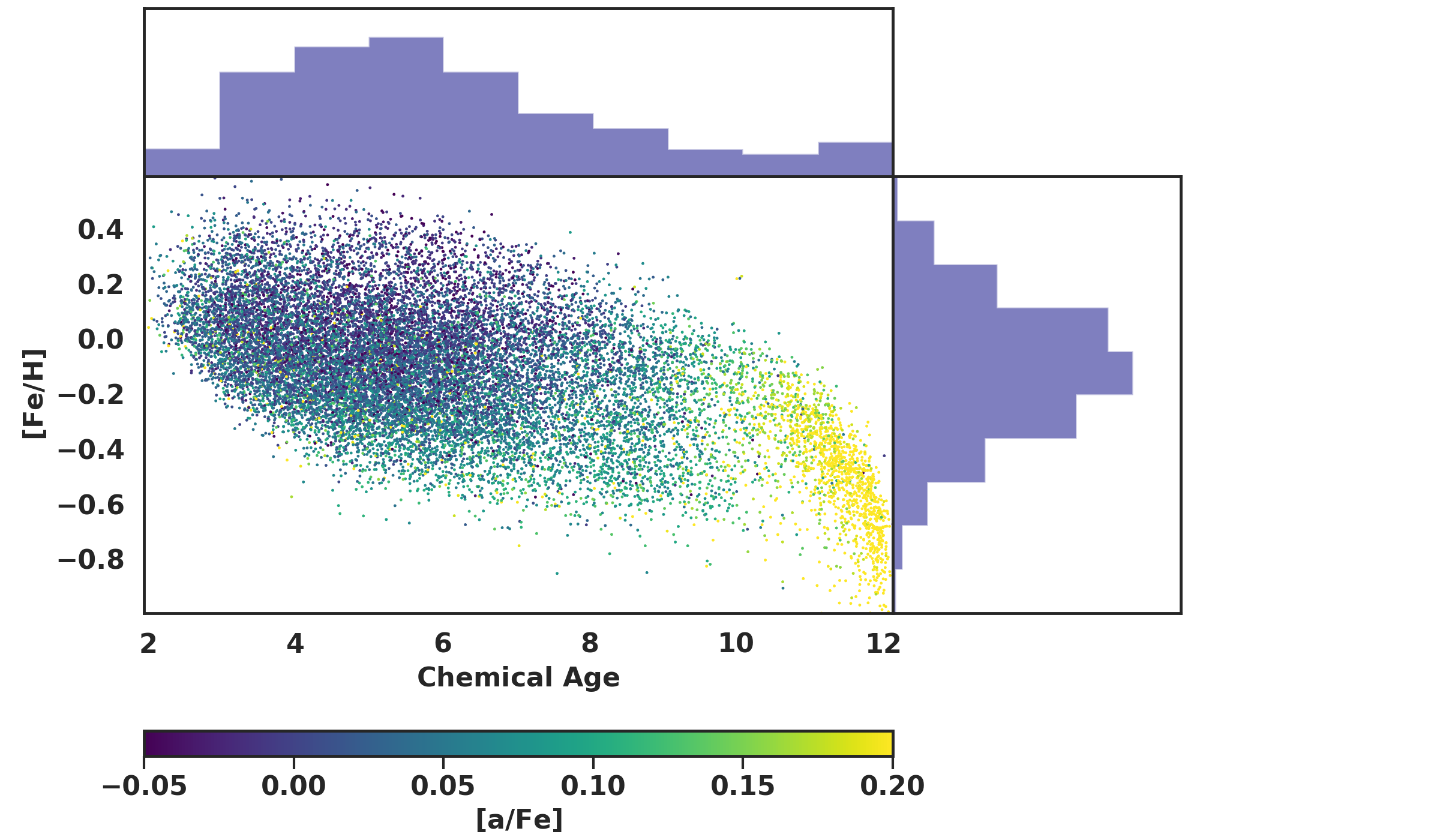}
\caption{\textbf{Left:} The age-metallicity relation for MSTO stars determined from isochrone ages. \textbf{Right:} The age-metallicity relation determined from chemical ages. }
\label{agemet}
\end{figure*}

\begin{figure*}
\centering
\includegraphics[width=0.99\textwidth]{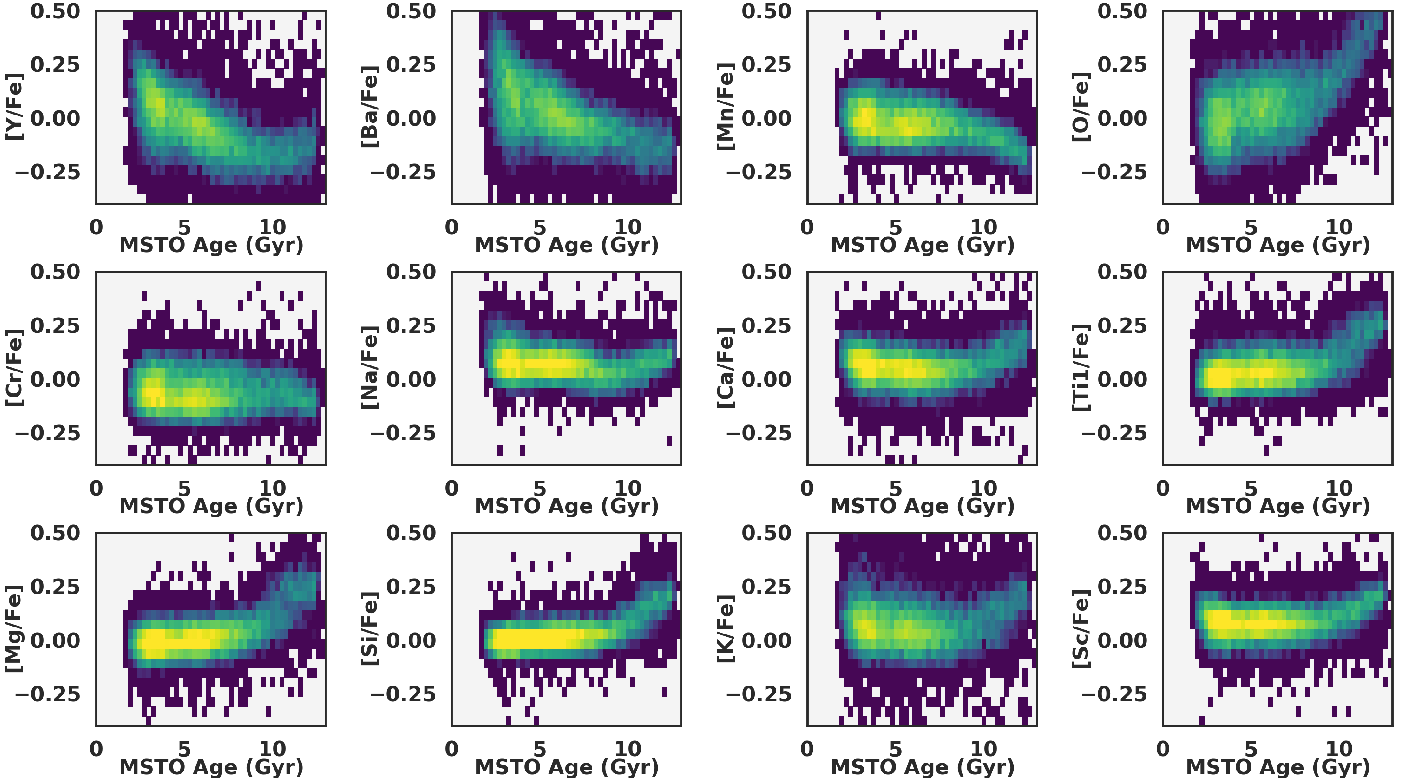}
\caption{The age-abundance trends for MSTO using isochrone ages.}
\label{mstoisoageabundance}
\end{figure*}

The relative importance of different abundances in determining the age using \xgb{} is shown in Fig. \ref{xgbgain}. The feature gain, which measures how useful each element is in improving the accuracy of the age estimates, is shown in the top panel. Perhaps unsurprisingly, the \afe{} elements [Mg/Fe] and [Si/Fe] are at the top, followed by the s-process element [Y/Fe]. This makes sense from a qualitative perspective: we know the oldest stars are enhanced in $\alpha$ elements, so we have an immediate age indicator just from the [Mg/Fe] and [Si/Fe] abundance if a star is (relatively) young or old. For younger and intermediate aged stars with lower [Mg/Fe] or [Si/Fe], \afe{} is less able to discriminate age. This is where the s-process elements become important in distinguishing young from intermediate aged populations. The bottom panel of Fig. \ref{xgbgain} is the relative weight of each element in \xgb. The weight describes how often an element occurs in different decision trees in \xgb. For this metric, the $\alpha$ elements are much lower down in the metric, with s-process, odd-Z, and iron peak elements being dominate, as they help distinguish between different locations within the Galaxy for a given age (i.e, for a given age there can be a range of metallicities which each correspond to a different birth radius). 

Uncertainties in the age estimates are calculated by doing 1,000 Monte Carlo runs of the input abundances through the model for each star, with the standard deviation of the age distribution being reported. The age distribution generated from the Monte Carlo run is generally a single peaked Gaussian. However, there is an exception for stars with ages between 8-10 Gyr. These stars typically have an age distribution that is bimodal, and the ages are quite sensitive to errors in \afe. Stars in this age range generally have thin disk/solar-\afe{} measurements, but still have very low s-process abundances, similar to that of the thick disk population. This means that $\alpha$ elements are the dominant age indicator distinguishing them from the thick disk. For these older thin disk stars, an error in \afe{} of 0.1 dex, for example, might shift a star towards the thick disk as the other most age-sensitive abundances are not yet reliable for age determination. 

\begin{figure*}
\centering
\includegraphics[trim={80 40 80 40},clip,width=0.99\textwidth]{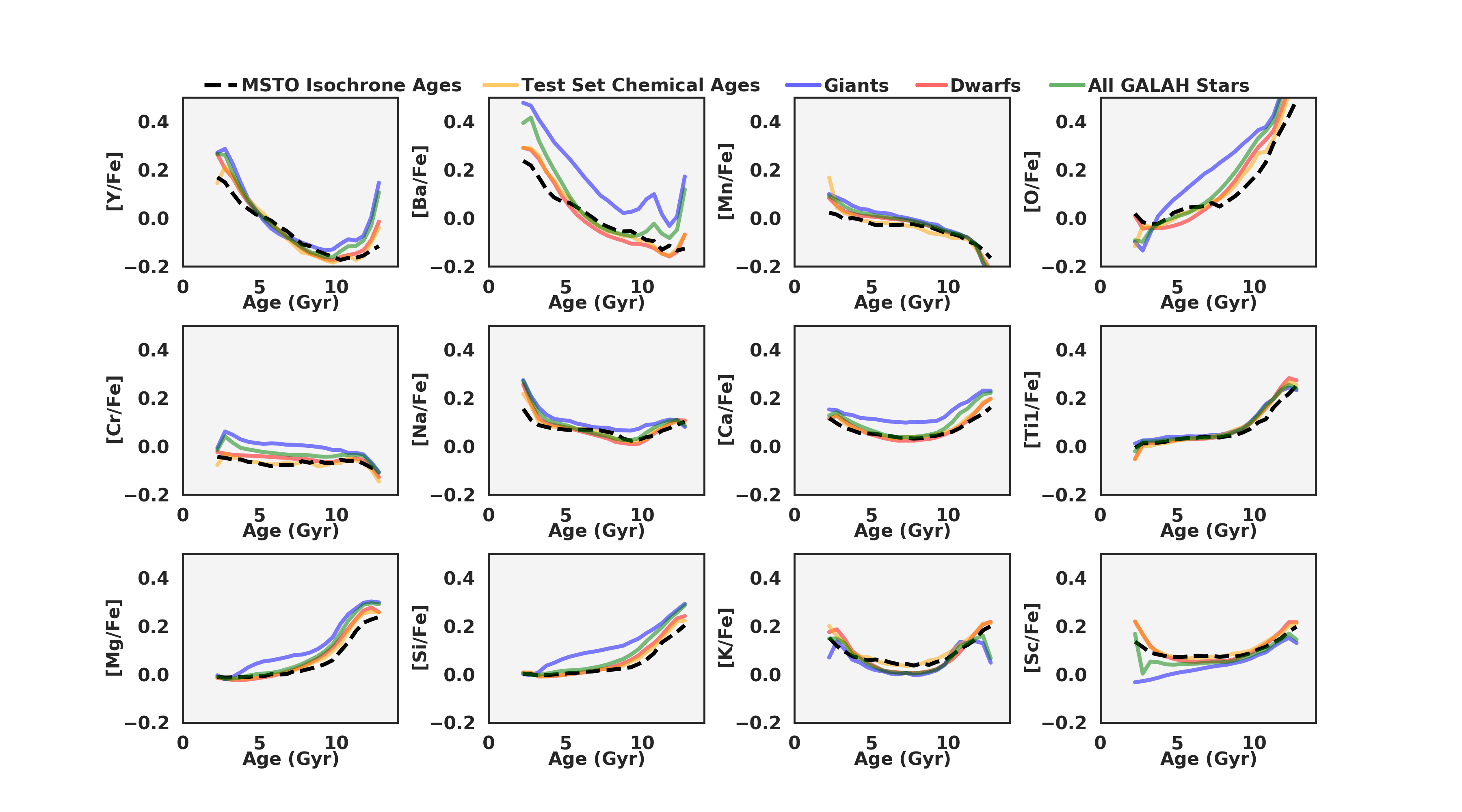}
\caption{The age-abundance trends for the various GALAH DR3 samples analyzed in this paper. Note the large discrepancy for the s-process elements [Y/Fe] and [Ba/Fe] for the giants.}
\label{allageabundance}
\end{figure*}

The median age uncertainty due to random errors as a function of SNR is shown in the top panel of Fig. \ref{ageuncertainty}. The median random errors for dwarfs and giants as a function of SNR are comparable, and between 0.5 and 1.5 Gyr, with the error in the age decreasing as signal to noise increases. The giants have a larger scatter as a function of SNR, however. We estimate the systematic errors by using the results of the MSTO test set. As we have calculated the random errors due to abundance uncertainties for the test set stars, the systematic errors inherent in the method can be estimated using the scatter about the relation found in Fig. \ref{testset}; this systematic uncertainty is ${\sim}1$ Gyr. The total uncertainty, taking both the random and systematic errors into account, is shown in the bottom panel of Fig. \ref{ageuncertainty}; typical age errors for each star are $\sim1-2$ Gyr. These results are more robust than those determined using Bayesian statistics, and we elect to use the ages derived from \xgb{} going forward. 

The age-metallicity relation for MSTO stars is shown in Fig. \ref{agemet}. The trends between the input training set from the isochrones is quite similar to the trends found using chemical ages; the age determined from chemistry having slightly less scatter, as expected given that the ages are determined directly from the chemistry in this case. The global features of the relations are present and in the same locations: the oldest stars are \afe{} enhanced, and there is a trend of increasing metallicity with decreasing age for these thick disk stars. Once $\tau\lesssim9$ Gyr, however, there is a distinct lack of an obvious age-metallicity relation in the thin disk, which mirrors results from previous studies (e.g., \citealt{casagrande2011,haywood2013,bergemann2014}).

\subsection{The Impact of Atomic Diffusion and Convective Mixing Length Variation}
The photospheric chemical abundances (i.e., those we estimate) can be different than the bulk chemical composition of a star due to atomic diffusion. Atomic diffusion is a catch all term to describe the various transport processes that are most efficient in the radiative zones of stars, driven by gradients in pressure, temperature, and concentration. Differences between the surface and bulk composition can lead to age errors on the order of 20\% for MSTO stars (see e.g., \citealt{thoul1994,dotter2017,liu2019}), becoming preferentially larger for older stars. Additionally, for MSTO stars in particular, atomic diffusion can cause abundance underestimates of $\sim0.1$ dex relative to the bulk composition. This means that atomic diffusion, if not taken into account, can cause an error in our calibration sample both in the age determination and in the chemical abundance relations used to derive the ages for the rest of the sample. The version of the PARSEC isochrones used in this analysis do take atomic diffusion into account, and we use the bulk composition in our age estimates, so the effect on age determination of MSTO stars is mitigated. Still, there remains the potential for diffusion to effect the abundances of our training set. The overall impact of this is likely small, given that the magnitude of the effect is only slightly larger than our abundance uncertainties, and most importantly because the abundance variation of [X/H] varies roughly in lock step with [Fe/H]. This means that the [X/Fe] ratio used in our analysis will be relatively unaffected, and the primary error in a chemical age determination will then be due to any [Fe/H] offset between the surface and bulk composition. This [Fe/H] offset could then yield a systematic error in age of up to 20\% in the chemical age of a star, but is in general much smaller than this value as [Fe/H] is not as important as $\alpha$ or s-process elemental abundances in the age estimations. 

Variations have been found in the convective mixing length both with metallicity (e.g., \citealt{tayar2017}) and there are also hints that there could be variations due to mass as well \citep{joyce2018b}. The primary impact of this variation in the convective mixing length is to cause an offset between the observed effective temperature of a star and the temperature of a star of that mass and metallicity in model isochrones, if the variation in mixing length is not accounted for in the models. As temperature is a critical component to the age estimation of MSTO stars, any offsets between observations and isochrones will therefore cause an error in the age determination. The impact on MSTO stars is largest for older (lower mass) and more metal-poor stars; \citet{viani2018} find for example a 0.3 difference in the mixing length parameter for an old metal-poor star relative to the solar mixing length, this causes a 200K shift towards cooler temperatures of the models on some portions of the turn off region. Similarly, in their study of metal-poor stars with [Fe/H]$\sim-2$, \citet{joyce2018a} find variations of the mixing length of 0.3, which also causes a shift of $sim150$K cooler along the MSTO region. This means that the ages would be underestimated compared to reality. However, the areas of the turnoff region which are most effected by this are relatively small (the turnoff itself only, see their Fig. 16), and the majority of the stars in our training set lie further along the MSTO or subgiant branch where the effective temperature difference between models is small. For the majority of our training set, our age determinations will be unaffected. For the older, more metal-poor stars, we might underestimate the age of an MSTO star by up to $\sim15$\% based on a temperature shift of 200K in the isochrones. 

As a whole, these effects are relatively minor compared to our global uncertainties, but do add to the systematic error budget. Particularly, these effects could be another possible explanation for the slight discrepancy between the chemical and isochrone ages for stars with $8<\tau<10$ Gyr. 

\section{Results}
\subsection{Age-Abundance Relations}
\begin{figure}
\centering
\includegraphics[trim={65 0 90 10},clip,width=0.48\textwidth]{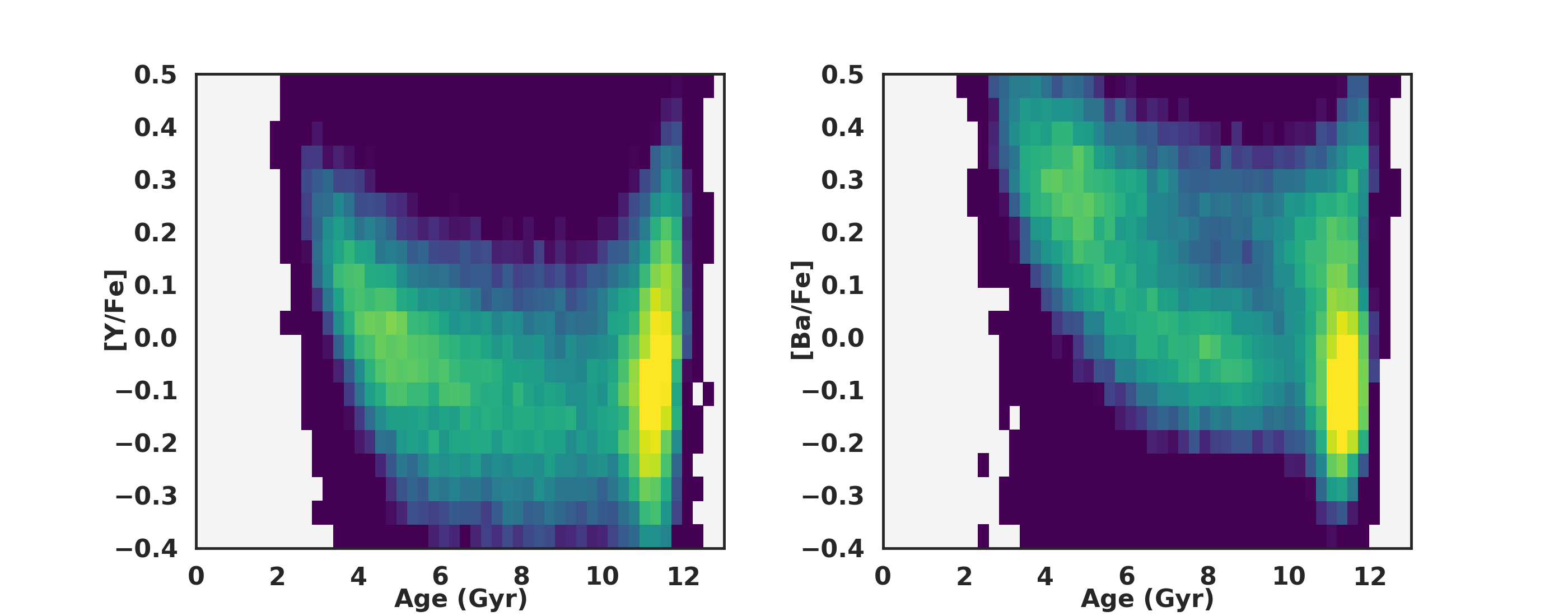}
\caption{The age-abundance trends of [Y/Fe] and [Ba/Fe] for giant stars. Note the large scatter in these s-process abundances for old stellar populations. This highlights a potential issue in the s-process abundance estimates for the old, metal-poor thick disk populations. }
\label{giantageabundance}
\end{figure}

Fig. 9 shows the age-abundance trends for the input training set. These are the same trends found in \citet{sharma2020c}. In the ideal case, the trends should match the training set sample exactly, as the chemical abundances themselves are being used to determine the ages. For the most part, we find that this is true, and the same abundance trends found in the training set are well matched to those for the entire sample derived with chemical ages, as shown in Fig. \ref{allageabundance}. These results closely match the age-abundance trends measured in \citet{sharma2020c}. However, there is a particular issue with the [Y/Fe] and [Ba/Fe] abundances for giants, where there is a strong discrepancy between the expected output and what is observed, with the [Y/Fe] and [Ba/Fe] abundances having large scatter for the oldest stellar populations which is not present in the MSTO or dwarf subsamples. The large scatter in [Y/Fe] and [Ba/Fe] for the giant subsample is shown in Fig. \ref{giantageabundance}.

\begin{figure}
\centering
\includegraphics[trim={40 45 0 0},width=0.49\textwidth]{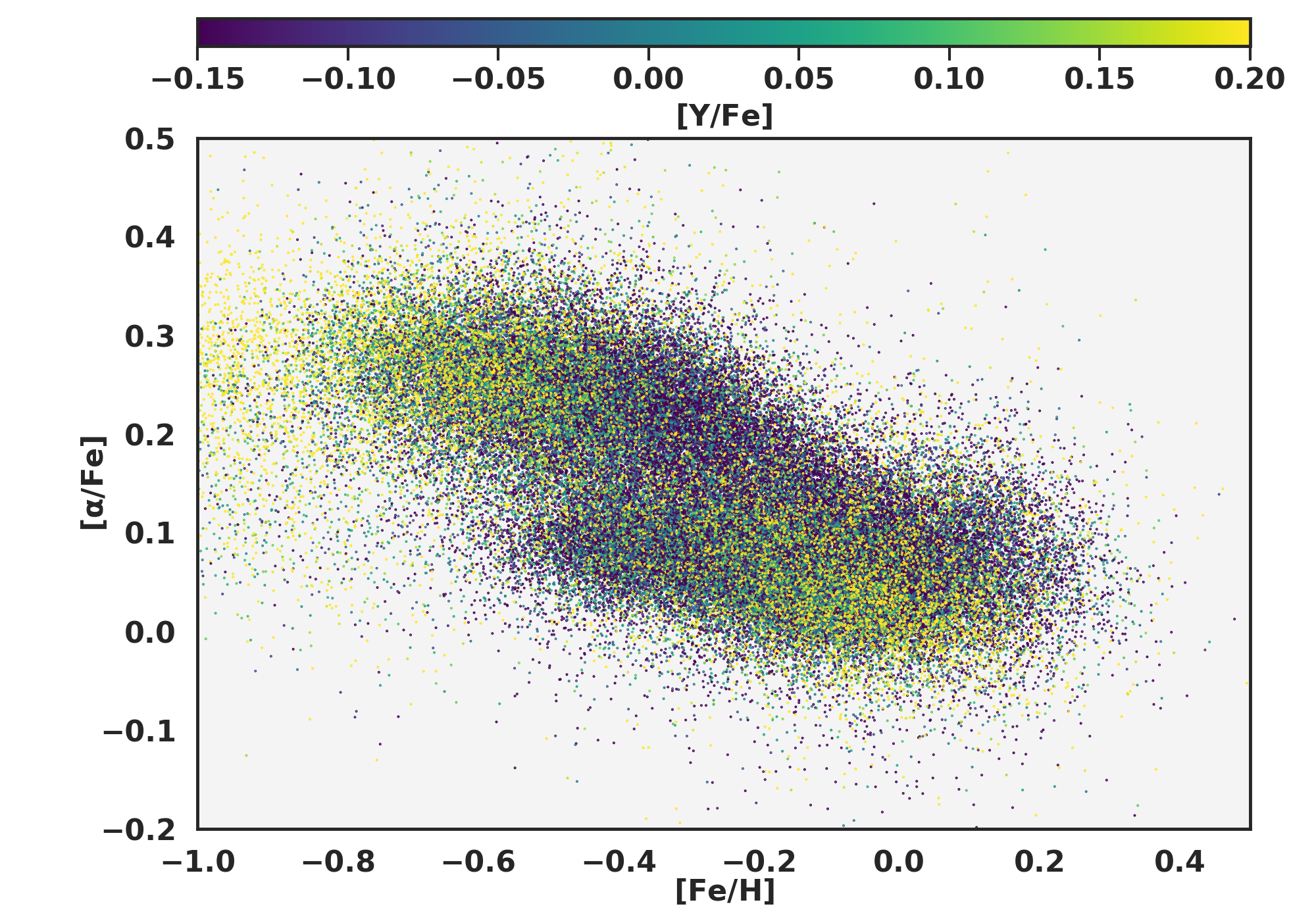}
\caption{The \afe{} vs. [Fe/H] plane for the giants. It is expected that young stars have higher [Y/Fe] than older stars, as [Y/Fe] is an s-process element. We identify a potential issue with the [Y/Fe] abundance for the metal-poor thick disk in the giant subsample ([Fe/H$<-0.5$, \afe$>0.15$), where many stars have enhanced [Y/Fe]. These stars should be among the oldest in the disk based on their belonging to the metal-poor thick disk.}
\label{giantsafefeh}
\end{figure}

The expectation for [Y/Fe] and [Ba/Fe], as s-process elements, is for low abundances at old ages, and high abundances at young ages, as is measured in the MSTO and dwarf samples; this trend is not reproduced in the giants highlighting an issue either with the abundance of [Y/Fe] and [Ba/Fe] or with the chemical age determination. We perform several tests to validate the reliability of the giant ages in light of the issue surrounding the s-process abundances for old stars. The \afe{} vs. [Fe/H] plane for the giant sample is shown in Fig. \ref{giantsafefeh}, with the color code being the [Y/Fe] abundance. The problem with the [Y/Fe] measurements for the oldest and most metal-poor stars is immediately apparent: there are a large fraction of stars with \afe{} and [Fe/H] belonging to the thick disk which have high [Y/Fe] measurements. These stars are likely some of the oldest in the Galaxy, based purely on their \afe{} and metallicity: this is the metal-poor end of the thick disk, so their [Y/Fe] should therefore be low. This hints at a problem in the [Y/Fe] and [Ba/Fe] abundances for metal-poor high-\afe{} stars. Luckily, for stars belonging to the thick disk, the [Y/Fe] and [Ba/Fe] are not useful tracers for age, as AGB stars have not produced large amounts of s-process elements, and the age can be derived almost entirely from the \afe{} and [Fe/H] abundances alone (see e.g., \citealt{haywood2013,hayden2017}).

\begin{figure}
\centering
\includegraphics[width=3.2in]{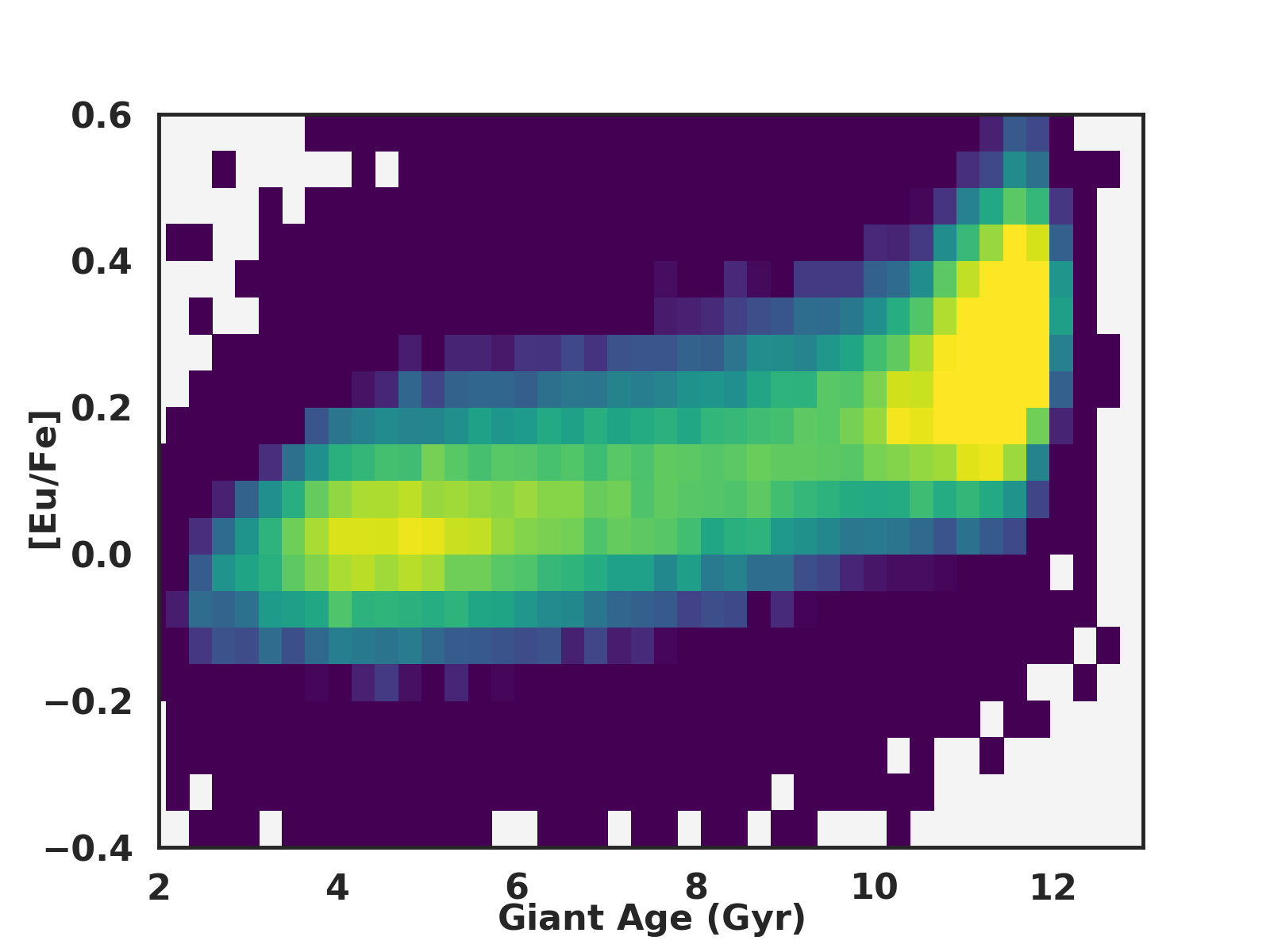}
\caption{The age-[Eu/Fe] trend for giants. [Eu/Fe] is an r-process element and should have a strong trend with age, which we recover in our analysis. [Eu/Fe] was not one of the elements used in the age generation. This means that the age determinations for giants are robust, despite the issue with the s-process abundances. }
\label{giantseu}
\end{figure}

The question of reliability for the [Y/Fe] and [Ba/Fe] and their impact on the age determination is then primarily for the thin disk, i.e., stars with age $<9$ Gyr. For these stars \afe{} is no longer a useful age tracer, and the [Y/Fe] and [Ba/Fe] become critically important for reliable age determination. For younger stars, the giant age-abundance trends match closely to those of the dwarfs and MSTO stars, which is an encouraging sign and hints that the [Y/Fe] and [Ba/Fe] are reliable where they are important for the age measurement. Further, we have additional abundance tracers that were not used in the age determinations available for giants for which to test the age reliability. As mentioned in the data section, [Eu/Fe] is not well measured in the MSTO or dwarf samples, but is measured in the majority of giants. As an r-process element, [Eu/Fe] is expected to have a strong age-abundance trend (e.g., \citealt{sharma2020c}). This makes [Eu/Fe] a useful diagnostic in our chemical age determinations, as we can measure the [Eu/Fe] vs. chemical age trend and compare this to expected trends from other studies. We find a strong trend of increasing [Eu/Fe] as a function of chemical age as shown in Fig. \ref{giantseu}, as expected for an r-process element. Our result closely matches the age-[Eu/Fe] trends measured in other studies (e.g., \citealt{sharma2020c}). This highlights that the ages derived for giants from chemical abundances are likely trustworthy, and the issue with the [Y/Fe] and [Ba/Fe] abundances only affects thick disk stars for which these abundances are not critically important in the age determination.

\subsection{Kinematic Trends}
\begin{figure}
\centering
\includegraphics[width=3.4in]{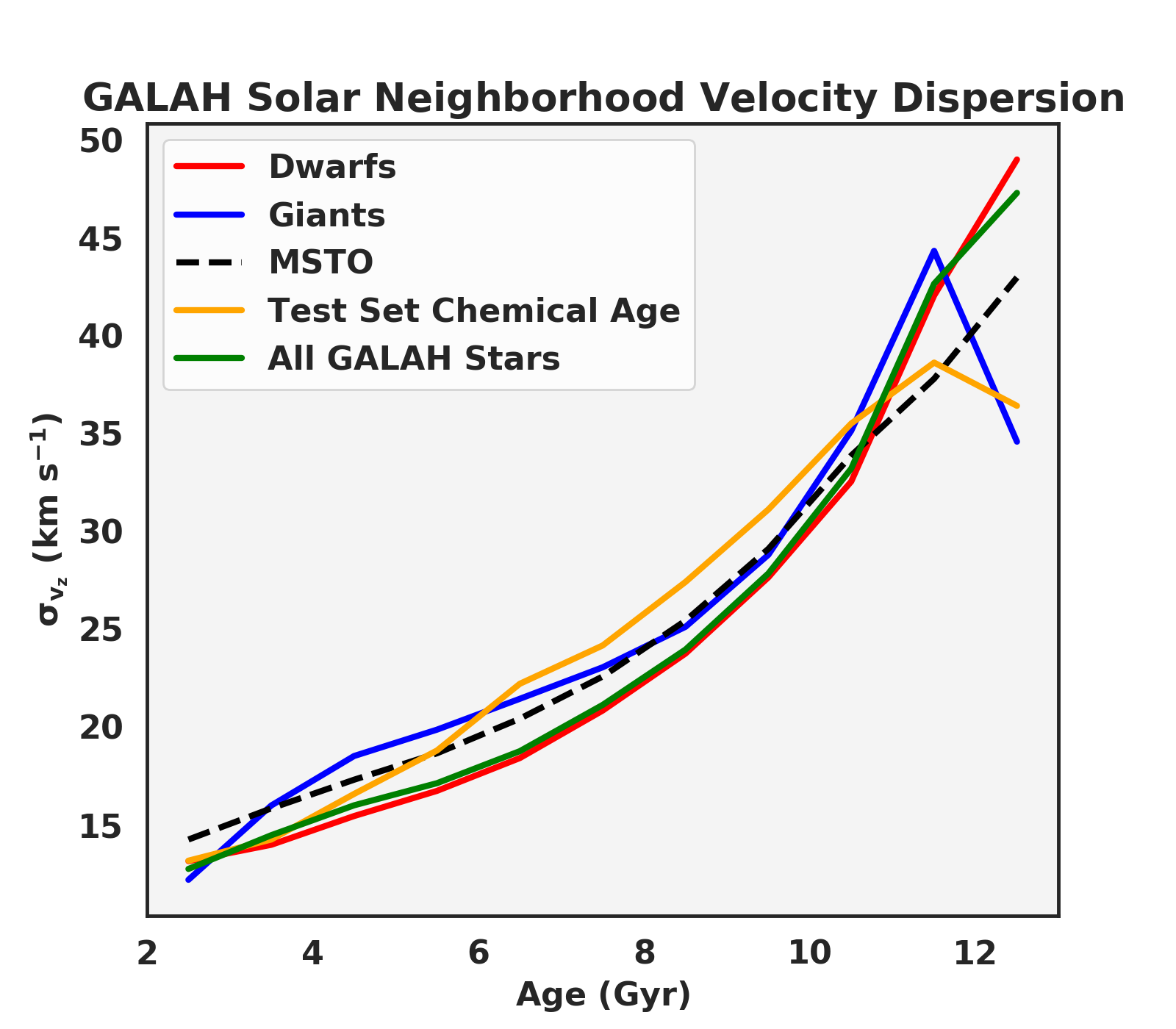}
\caption{The vertical velocity dispersion for different samples as a function of age in the solar neighborhood ($7.1<R<9.1$ kpc, $|z|<0.5$ kpc)}
\label{vdspsn}
\end{figure}

We demonstrate both the utility and accuracy of ages derived from chemical abundances by reproducing some of the recent results on the kinematic structure of the disk (e.g., \citealt{mackereth2019,sharma2020a}). These results also highlight that the potential issue with the [Y/Fe] and [Ba/Fe] abundances for giants belonging to the metal-poor thick disk does not dramatically impact the age determination, as the recovered kinematic relations generally agree for both giants and dwarfs. As shown in Fig. \ref{vdspsn}, we reproduce the age-velocity dispersion relation for the solar neighborhood ($7.1< R < 9.1$ kpc, $|z|<0.5$ kpc), and find that all of the subsamples agree well to within the errors in both the measured ages and the velocities. This result also closely matches previous studies of the vertical velocity dispersion in the solar neighborhood, who find a low velocity dispersion of $\sim10-15$ \vel{} for younger stellar populations that increases with age up to values between 40-50 \vel{} for the oldest stars in the disk \citep{nordstrom2004,haywood2013,feuillet2016,hayden2017,yu2018,sharma2020a}. Some authors have found a step function in the velocity dispersion between low- and high-\afe{} populations (e.g., \citealt{silvaaguirre2018,miglio2020}), but this is likely due to the definition of low- and high-\afe{} and imposing a hard \afe{} cut on the boundary between thin and thick disks, when the chemical distributions are actually rather continuous (see \citealt{sharma2020b}). We find that the velocity dispersion increases smoothly with age and that there is no step function in the estimated dispersion, although there is a clear inflection point around $\sim8-9$ Gyr, with a steeper increase in dispersion with age for older stellar populations.

We also apply the data set to cover a larger fraction of the Galaxy, and reproduce the results shown in \citet{sharma2020a}. \citet{sharma2020a} measure how the velocity dispersion of the disk varies with age, angular momentum, [Fe/H], and height about the plane ($|z|$). We apply the same formalism used in their paper to disentangle how the velocity dispersion depends on these parameters, as outlined below. The dispersion $\sigma_{\varv}$ of velocity $\varv$ (for either $\varv_R$ or $\varv_z$), is assumed to depend on the stellar 
age $\tau$, angular momentum $L_z$, metallicity [Fe/H], and vertical height from the disc midplane $z$, via the following multiplicatively separable functional form:
\begin{equation}
\sigma_{\varv}(X,\theta_{\varv})=\sigma_{\varv}(\tau,L_z,{\rm [Fe/H]},z,\theta_{\varv}) = \sigma_{0,\varv} f_{\tau}f_{L_z}f_{{\rm [Fe/H]}} f_{z}.
\label{equ:vdisp_model}
\end{equation}
Here, $X=\{\tau,L_z,{\rm [Fe/H]},z\}$ is a set of observables that are 
independent variables and 
\begin{equation}
f_{\tau}=\left(\frac{\tau/{\rm Gyr}+0.1}{10+0.1}\right)^{\beta_{\varv}},
\label{equ:f_tau}
\end{equation}
\begin{equation}
f_{L_z}=\frac{\alpha_{L,\varv} (L_z/L_{z,\odot})^2+\exp[-(L_z-L_{z,\odot})/\lambda_{L,\varv}]}{1+\alpha_{L,\varv}},
\label{equ:f_lz}
\end{equation}
\begin{equation}
f_{\rm [Fe/H]}=1+\gamma_{{\rm [Fe/H]},\varv} {\rm [Fe/H]},
\label{equ:f_feh}
\end{equation}
\begin{equation}
f_{z}=1+\gamma_{z,\varv} |z|,
\label{equ:f_z}
\end{equation}
and $\theta_{\varv}=\{\sigma_{0,\varv},\beta_{\varv},\lambda_{L,\varv},\alpha_{L,\varv},\gamma_{{\rm [Fe/H]},\varv},\gamma_{z,\varv}\}$ is a set of free parameters. We assume the same values used in their paper outlined in their Table 2 for these free parameters.

\begin{figure*}
\centering
\includegraphics[trim={140 0 160 10},clip,width=0.99\textwidth]{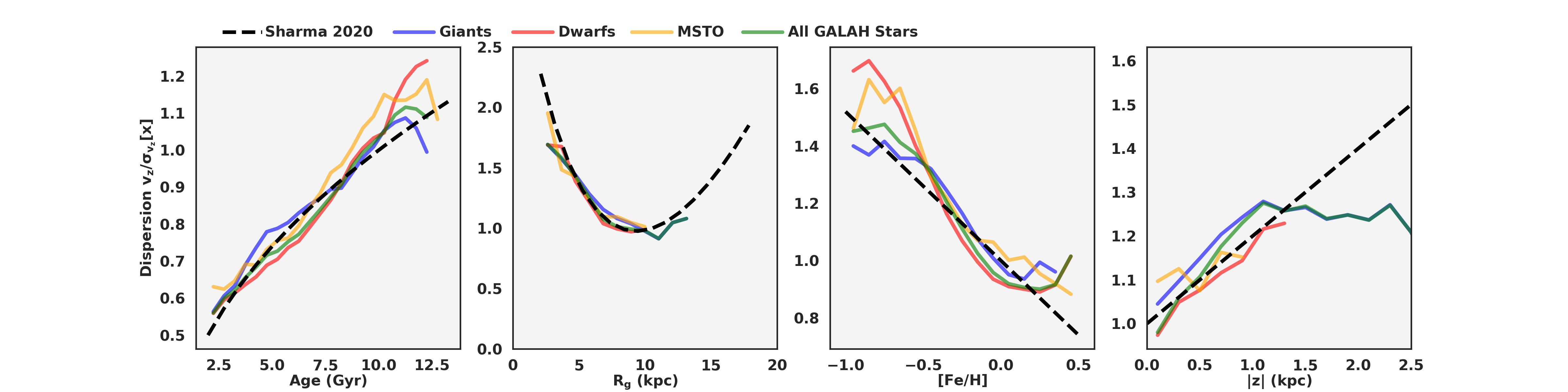}
\caption{The vertical velocity dispersion for different samples as a function of age, angular momentum, [Fe/H], and height above the plane. Note that the test set relation is omitted in this figure due to insufficient sampling.}
\label{vdspsharma}
\end{figure*}

Our results, shown in Fig. \ref{vdspsharma}, match closely with the results of \citet{sharma2020a} (Fig. 1 in their paper) for all of our samples. Briefly, we find that the velocity dispersion increases with age following a shallow power law slope. The velocity dispersion as a function of the angular momentum, or guiding radius, is high in the inner Galaxy, while being lower in the solar neighborhood with a flattening, or perhaps hints of an increase, in the outer disk. As GALAH is a southern hemisphere survey, it is difficult to probe the outer Galaxy as compared to the analysis in \citet{sharma2020a}, who also utilized observations from LAMOST \citep{deng2012} to determine the relation in the outer disk. The dependence on metallicity is roughly linear, although there is an S shape present in the dispersions for the dwarf stars relative to the model; this is still similar to what is observed by \citet{sharma2020a}. The dependence on metallicity is the only panel that shows any difference on the velocity dispersion relation between the various subsamples characterized in this paper, with the giants following the linear trend more closely, but even here the difference is relatively minor between dwarfs and giants. The only major difference in our results compared to \citet{sharma2020a} is the vastly increased coverage of the thick disk in the sample presented in this paper, showing that while a linear dependence on $|z|$ is a good fit for stars up to $|z|<1.5$ kpc, it may not hold for stars high above the plane. The vastly increased sample size in our analysis, in particular for the thick disk, is made possible by the use of chemical ages, rather than being forced to rely on MSTO or \textit{Kepler} asteroseismic values for age estimates which have more limited Galactic coverage.

\subsection{Catalog and Access}
This catalog will be available online through MNRAS, as well as locally at the University of Sydney here: \url{ http://physics.usyd.edu.au/~mhay5097/chemicalclocks/} once the paper is accepted. The catalog will contain the sobject\_id for a given star, the age estimate for that star, and the total uncertainty in the age estimate.
%

\section{Discussion}
We calculate the ages of stars in GALAH DR3 using their measured chemical abundances. With these chemical ages, we are able to reproduce the global kinematic properties of the disk, such as the age-angular momentum-velocity dispersion relation found by \citet{sharma2020a}. We have demonstrated that ages can be accurately predicted for stars with a large range of birth radii, metallicity, and chemical abundances. These results, as well as those in \citet{ness2019} and \citet{sharma2020c}, demonstrate that the age, metallicity, and overall chemical abundance do not vary dramatically at a given birth radius.

The results presented here, as well as those in \citet{ness2019} and \citet{sharma2020c} highlight that the variation in abundances at a given place in time in the Galaxy is quite small. This means that strong chemical tagging, particularly for stars belonging to the thin disk, may prove to be difficult with our current abundance precision. The deviations from age-abundance trends measured by \citet{ness2019} are roughly the same size as the abundance measurement uncertainties, and \citet{sharma2020c} find similar results using GALAH observations covering a wider range of chemical abundances and Galactic spatial coverage. To first order, this means that the dominant factor in the abundance of a star is its age and birth radius. Early strong chemical tagging efforts find many groups in chemical space, but have been unable to recover open clusters observed in major spectroscopic surveys relative to the background disk stars (e.g., \citealt{pricejones2020}). Verification of these groups is difficult, as our results demonstrate that by having similar chemical abundances, stars grouped in chemical space will by necessity have similar age. This means that having a similar age or falling along an isochrone cannot be used as validation for strong chemical tagging, as stars will by definition meet this criteria simply by having similar chemical abundances. Until it is demonstrated that open clusters can be chemically tagged reliably, claims of strong chemical tagging should be viewed with some scepticism, given our results and the arguments laid out in \citet{ness2019} and \citet{sharma2020c}.  

However, it is possible that future improvements in the precision of abundance determination, as well as including additional elements, could make strong chemical tagging more tangible going forward. For instance, \citet{2010ApJ...713..166B} estimate that abundance precision of 0.01 to 0.02 dex is needed to enable reliable strong chemical tagging. This is still likely lower than our current uncertainties, but we are approaching this value, particularly with data driven machine learning methods such as the Cannon \citep{ness2015} or Astro-NN \citep{leung2019}. Continued improvements in abundance determination, such as improved corrections for non-local thermal equilibrium effects (see e.g., \citealt{amarsi2020}) as well as better reduction techniques to eliminate issues such as fiber cross-talk \citep{kos2018}, could mean that we are close to the required abundance precision necessary to enable strong chemical tagging.

For the Milky Way, the fact that abundances can be determined merely by a stars age and metallicity \citep{ness2019,sharma2020c}, or conversely accurate ages directly from abundances, likely means that the gas azimuthal mixing timescales are quite short. Strong azimuthal abundance variation would yield a much larger scatter in these relations and hints that the Milky Way must be fairly uniform with azimuth. However, recent observations of HII regions in the Milky Way make claims of azimuthal metallicity variation across the Galaxy \citep{wenger2019}. The results from the HII regions are difficult to mesh with the conclusions of \citet{ness2019}, \citet{sharma2020c}, and this paper; it is possible that the results coming from HII regions is due to large uncertainties in the distances to Galactic HII regions. Observations of azimuthal abundance variation in external spiral galaxies are inconclusive: some galaxies have azimuthal variation, often correlating with metallicity peaks along the spiral arms or due to interactions with the bar, while other galaxies with seemingly similar properties do not (e.g., \citealt{kreckel2019}). This presents a somewhat muddled view of azimuthal abundance variations and it is unclear how the Milky Way fits into this picture, other than that at least for most of its history the Milky Way seems to have been fairly well mixed azimuthally, as otherwise the age-abundance relations and the chemical age determinations would not work as well as they do.

While strong chemical tagging may prove difficult with current spectroscopic studies, this is not necessarily bad news for Galactic Archaeology. The results presented in this paper highlight the capability of weak chemical tagging which enable a host of new opportunities and pathways for studying the Milky Way, and in particular the kinematic structure and evolution of the Galaxy. The fact that we can use chemistry alone to estimate a robust age for a star has important implications for future studies of the Milky Way. Previously, kinematic studies were often limited to MSTO stars or asteroseismic targets \citep{sharma2020a}, or use chemistry where ages were not available (e.g., \citealt{hayden2018,hayden2020}). The ability to use abundances to estimate a reliable age provides not only a greatly increased sample size, but also vastly increases the Galactic coverage. MSTOs probe a limited volume given the combination of their absolute magnitudes and the magnitude limits imposed by spectroscopic surveys; for example, in GALAH MSTO stars probe a volume of $\sim1.5$ kpc around the solar position. Relying on asteroseismic targets for ages restricts you to the \textit{Kepler} and \textit{K2} fields, or the bright magnitude limits of the \textit{TESS} mission. Additionally, asteroseismic ages also have much larger uncertainties than those in MSTO stars, and potentially the ages presented in this paper. For example, the age-\afe{} relation found in studies of asteroseismic giants has significantly more scatter than those of MSTO stars (see e.g., \citealt{silvaaguirre2018,miglio2020}).

Applying the methods outlined in this paper is also possible for other completed, ongoing, and future spectroscopic surveys. In particular, the existing datasets of LAMOST and APOGEE and the upcoming surveys 4-MOST \citep{dejong2014} and WEAVE \citep{dalton2014} are prime targets for such an analysis. In the ideal case, putting these surveys on the same abundance scale (and, by the methods outlined in this paper, the same age scale) would enable them to be used together, providing a sample of several million stars with reliable ages determined directly from chemical abundances. A data set of this magnitude would allow for precise constraints on important kinematic properties that govern the secular evolution of the disk such as the efficiency and timescales of blurring and migration. It would also allow, for example, a detailed study of the age structure of the ridges observed in $V_{\phi}$-$R$ space \citep{antoja2018,khanna2019}, the phase-spiral \citep{antoja2018,bland-hawthorn2019}, or galactoseismology \citep{widrow2012,bland-hawthorn2020}.

The results presented here also highlight the need for future spectroscopic surveys to cover as many nucleosynthetic channels as possible, while also providing high quality and high signal to noise ratio spectra. As clearly seen in Fig. \ref{ageuncertainty}, the age precision derived from chemical abundances is a direct function of the signal to noise ratio of the spectra. Higher quality observations will lead to a direct improvement in the age determinations. The need for precision abundances is also highlighted in \citet{sharma2020c}: while the age-abundance trend for every element is dominated by the age and metallicity, there are still smaller second order effects with signal to noise, \teff, and \logg. There is also the potential for Galactic evolution effects on smaller scales, such as the offsets in the [Y/Mg] relations as a function of metallicity identified by \citet{casali2020}; these effects will be difficult to detect if other sources of uncertainty in the age-abundance relation dominate due to low quality observations. Additional elements, particularly those coming from the s-process, will also provide tighter constraints on the age determination as well. 

\section{Conclusions}
In this paper we have determined ages for a large fraction of the stars in the GALAH survey with well-measured chemical abundances, using the \xgb{} algorithm. The training set consists of high SNR MSTO stars with precision ages determined from isochrone matching. The catalog is publicly available and ideal for kinematic studies of the Galaxy. We demonstrate that the stellar ages derived from chemical abundances are reliable and can reproduce many of the age-kinematic results that have been observed since the advent of \gai. In particular, we reproduce the age-velocity dispersion of the solar neighbourhood, and of the global relation found by \citet{sharma2020c}, but with a greatly increased GALAH sample and much improved coverage of the thick disk. This form of weak chemical tagging, which enables an order of magnitude increase in the sample size as well as increasing the sample volume of stars which have reliable age estimates, enables more detailed studies of the kinematic structure of the disk. 

When combined with the results from \citet{ness2019} and \citet{sharma2020c}, strong chemical tagging seems unlikely with our current abundance precision of $\sim0.05$ dex. The ability to measure precise ages and recreate the age-kinematic trends across the disk, as well as find very small dispersions in age-abundance relations, requires that the chemical abundance variation at a given radius and time is small. This is a boon for weak chemical tagging and the study of the kinematic structure of the Galaxy, but means that our current abundance precision is likely not high enough to engage in strong chemical tagging.

Going forward, the methods outlined in this paper can be applied to several of the existing and upcoming Galactic Archaeology surveys. If these surveys can be put on the same metallicity scale, reliable ages can be determined for millions of stars while also having excellent coverage of the Galaxy spatially. This dataset would be truly unprecedented, and when combined with \gai{} DR3, our understanding of the formation and evolution of the Milky Way could undergo another gigantic leap forward. 

\section{Acknowledgements} This work is also based
on data acquired from the Australian Astronomical Telescope.
We acknowledge the traditional owners of the land
on which the AAT stands, the Gamilaraay people, and pay
our respects to elders past and present. This research was supported by the Australian Research Council Centre of Excellence for All Sky Astrophysics in 3 Dimensions (ASTRO 3D), through project number CE170100013. This work has made use of data from the European Space
Agency (ESA) mission \gai (https://www.cosmos.esa.
int/gaia), processed by the \gai Data Processing and
Analysis Consortium (DPAC, https://www.cosmos.esa.
int/web/gaia/dpac/consortium).

In addition to ASTRO3D, MRH received support from ARC DP grant DP160103747. LS acknowledges financial support from the Australian Research Council (discovery Project 170100521). JK and TZ acknowledge financial support of the Slovenian Research Agency (research core funding No. P1-0188) and the European Space Agency (PRODEX Experiment Arrangement No. C4000127986). KL acknowledges funds from the European Research Council (ERC) under the European Union's Horizon 2020 research and innovation programme (Grant agreement No. 852977). SLM and JDS acknowledge support from the UNSW Scientia Fellowship program and the Australian Research Council through grant DP180101791. YST is grateful to be supported by the NASA Hubble Fellowship grant HST-HF2-51425.001 awarded by the Space Telescope Science Institute.

\bibliographystyle{mnras}
\bibliography{ref}
\end{document}